\documentclass[twocolumn,tighten]{aastex63}
\usepackage{color}
\definecolor{mygreen}{rgb}{0,0.6,0}
\definecolor{mygray}{rgb}{0.5,0.5,0.5}
\definecolor{mydarkgray}{rgb}{0.3,0.3,0.3}
\definecolor{myred}{rgb}{0.8,0,0}
\definecolor{myblue}{rgb}{0.0,0,0.9}

\usepackage{amsmath}	
\usepackage{bm}	
\usepackage{enumitem}
\usepackage{listings}

\setcitestyle{notesep={; }}

\hypersetup{
  bookmarks=false,         
  bookmarksopen=false,     
  pdffitwindow=true,       
  pdftitle={Hypercubes of AGN Tori (HYPERCAT) -- I. Models and Image Morphology},    
  pdfauthor={R. Nikutta et al.},   
  colorlinks=true,         
  linkcolor=blue,          
  citecolor=blue,          
  urlcolor=blue,           
  linktocpage=true
}

\def\about         {\hbox{$\sim$}}
\renewcommand\deg  {\ensuremath{\rm deg}}
\newcommand\dif    {\hbox{${d}$}}
\newcommand\etapix {\ensuremath{\eta}}

\newcommand\Lb   {\ensuremath{L_{\rm bol}}}
\newcommand\mic  {\hbox{$\mu{\rm m}$}}

\newcommand\C {\textsc{Clumpy}}        
\newcommand\D {\textsc{Dusty}}
\newcommand\HC {\textsc{Hypercat}}        

\newcommand\xax  {\hbox{$x$-axis}}
\newcommand\yax  {\hbox{$y$-axis}}
\newcommand\zax  {\hbox{$z$-axis}}
\newcommand\N    {\hbox{${\cal N}$}}  
\newcommand\Nc   {\ensuremath{N_{C}}}
\newcommand\No   {\hbox{${\cal N}_0$}}    
\newcommand\NT   {\hbox{${\cal N}_{T}$}}  
\newcommand\iv  {\hbox{$i$}} 
\newcommand\sig  {\hbox{$\sigma$}}
\newcommand\q    {\hbox{$q$}}
\newcommand\tv   {\hbox{$\tau_{V}$}}  
\newcommand\Y    {\hbox{$Y$}}
\newcommand\Ymax {\hbox{$Y_{\rm max}$}}
\newcommand\pa   {\ensuremath{{\rm PA}}}
\newcommand\Rd   {\ensuremath{R_{d}}}
\newcommand\RH {\ensuremath{R_{\rm HL}}}
\newcommand\RHorig {\ensuremath{R_{\rm HL}^{\rm origin}}}

\newcommand\Rgx {\ensuremath{R_g^x}}
\newcommand\Rgy {\ensuremath{R_g^y}}
\newcommand\Sc   {\ensuremath{S_{C,\lambda}}}

\newcommand\ALMA {ALMA}
\newcommand\ELT {ELT}
\newcommand\GMT {GMT}
\newcommand\JWST {JWST}
\newcommand\Keck {Keck}
\newcommand\TMT {TMT}
\newcommand\VLTI {VLTI}

\defcitealias{Nenkova+2008a}{N08a}
\def\nenka {\citetalias{Nenkova+2008a}}

\defcitealias{Nenkova+2008b}{N08b}
\def\nenkb {\citetalias{Nenkova+2008b}}

\defcitealias{gravity2020}{G20}
\def\gravity {\citetalias{gravity2020}}

\defcitealias{Lopez-Rodriguez+2018}{LR18}
\def\loro {\citetalias{Lopez-Rodriguez+2018}}

\defcitealias{hc-paper2}{Paper II}
\def\papertwo {\citetalias{hc-paper2}}

\shorttitle{Hypercat -- I. Models and Image Morphology}
\shortauthors{Nikutta et al.}

\begin{document}
\title[HYPERCAT -- I. Models and Image Morphology]{Hypercubes of AGN Tori (HYPERCAT) -- I. Models and Image Morphology}



\correspondingauthor{Robert Nikutta}
\email{robert.nikutta@noirlab.edu}

\author[0000-0002-7052-6900]{Robert Nikutta} 
\affiliation{NSF's NOIRLab, 950 N. Cherry Ave., Tucson, AZ 85719, USA}

\author[0000-0001-5357-6538]{Enrique Lopez-Rodriguez}
\affiliation{Kavli Institute for Particle Astrophysics and Cosmology (KIPAC), Stanford University, Stanford, CA 94305, USA}

\author[0000-0002-4377-903X]{Kohei Ichikawa}
\affiliation{Frontier Research Institute for Interdisciplinary Sciences, Tohoku University, Sendai 980-8578, Japan}
\affiliation{Astronomical Institute, Tohoku University, Aramaki, Aoba-ku, Sendai, Miyagi 980-8578, Japan}

\author[0000-0003-4209-639X]{N. A. Levenson}
\affiliation{Space Telescope Science Institute, Baltimore, MD 21218, USA}

\author[0000-0001-7827-5758]{Christopher Packham}
\affiliation{Department of Physics \& Astronomy, University of Texas at San Antonio, One UTSA Circle, San Antonio, TX 78249, USA}
\affiliation{National Astronomical Observatory of Japan, 2-21-1 Osawa, Mitaka, Tokyo 181-8588, Japan}

\author[0000-0002-6353-1111]{Sebastian F. H\"{o}nig}
\affiliation{School of Physics \& Astronomy, University of Southampton, Southampton SO17 1BJ, United Kingdom}

\author[0000-0001-6794-2519]{Almudena Alonso-Herrero}
\affiliation{Centro de Astrobiolog\'{\i}a (CAB, CSIC-INTA), ESAC Campus, Villanueva de la Ca\~nada, E-28692 Madrid, Spain}

\begin{abstract}
\noindent
  Near- and mid-infrared interferometers have resolved the dusty
  parsec-scale obscurer (torus) around nearby active galactic nuclei
  (AGNs). With the arrival of extremely large single-aperture
  telescopes, the emission morphology will soon be resolvable
  unambiguously, without modeling directly the underlying brightness
  distribution probed by interferometers today. Simulations must
  instead deliver the projected 2D brightness distribution as a result
  of radiative transfer through a 3D distribution of dusty matter
  around the AGN.
  We employ such physically motivated 3D dust distributions in tori
  around AGNs to compute 2D images of the emergent thermal emission,
  using \C, a dust radiative transfer code for clumpy media.
  We demonstrate that \C\ models can exhibit morphologies with
  significant polar elongation in the mid-infrared (i.e. the emission
  extends perpendicular to the dust distribution) on scales of several
  parsecs, in line with observations in several nearby AGNs. We
  characterize the emission and cloud distribution morphologies. The
  observed emission from near- to mid-infrared wavelengths generally
  does not trace the bulk of the cloud distribution. The elongation of
  the emission is sensitive to the torus opening angle or scale
  height. For cloud distributions with a flat radial profile, polar
  extended emission is realized only at wavelengths shorter than
  $\about18\,\mic$, and shorter than $\about 5~\mic$ for steep
  profiles.
  We make the full results available through \HC, a large hypercube of
  resolved AGN torus brightness maps computed with \C. \HC\ also
  comprises software to process and analyze such large data cubes and
  provides tools to simulate observations with various current and
  future telescopes.

\end{abstract}


%
\section{Introduction}
\setcounter{footnote}{9}
\noindent
Unification of active galactic nuclei \citep[AGNs;
e.g.,][]{Antonucci1993, UrryPadovani1995} has enjoyed success for three
decades. It explained most of the observed dichotomy between type~1
and type~2 AGNs with the simple assumption of an axially symmetric,
anisotropic, dusty obscurer -- a torus -- and its orientation to the
observer's line of sight (LOS). AGN infrared (IR) spectral energy
distributions (SEDs) were initially modeled as emission from
smooth-density dusty tori (e.g., \citealt{PierKrolik1992,
  PierKrolik1993, GranatoDanese1994,
  EfstathiouRowan-Robinson1995}). However, the dust must be confined
in discrete clouds lest the torus be dynamically destroyed very
quickly \citep{KrolikBegelman1988}. In the X-ray regime symmetric
variability of the obscuring column density, $N_{\rm H}$, suggests a
circumnuclear medium composed of discrete gas and dust clouds;
individual clouds are witnessed crossing the LOS while orbiting well
inside the dusty parts of the torus \citep*[][and references
therein]{MKN2014}.

Radiative transfer models of clumpy circumnuclear dust are now
numerous (e.g., \citealt{Nenkova+2002}, \citeyear{Nenkova+2008a}, \citeyear{Nenkova+2008b},
hereafter N08a, N08b; \citealt{Hoenig+2006, Schartmann+2008,
  Stalevski+2012, Siebenmorgen+2015}). They have all been extensively
mined for their resulting SEDs. Studies of the spatially resolved
emission morphology, on the other hand, have so far been
underappreciated, presumably because resolved observations are only
recently possible. Resolved imagery is the only remedy for model
degeneracies that are inherent to IR radiative transfer, which is when
different dust geometries produce similar (unresolved) SEDs
\citep{Vinkovic+2003}. Before the dawn of 30~m class telescopes, only
interferometry has the potential to spatially resolve subparsec
mid-IR (MIR) emission around the nucleus. The VLTI/MIDI
instrument has resolved several nearby bright AGNs at MIR wavelengths
\citep[e.g.,][]{Jaffe+2004, Poncelet+2006, Tristram+2007, Raban+2009,
  Burtscher+2013, Tristram+2014, Leftley+2018}.

MIDI combines the light of only two telescopes; thus, phase closure
cannot be achieved, and images cannot be reconstructed without model
assumptions. Most authors resorted to modeling directly with a
combination of 2D Gaussian components in the plane of the sky the
brightness distribution that \VLTI\ has sampled. The semiaxes of the
Gaussians were varied, as were their orientations, the effective
blackbody temperatures generating each Gaussian, and the optical depth
and dust properties along the LOS toward each component. A
well-fitting model such as the one applied to \VLTI/MIDI observations
of Circinus \citep{Tristram+2014} required no fewer than 18 free
parameters for three Gaussian components. These empirical models do
not afford unambiguous conclusions about the underlying 3D dust
distribution.

At longer wavelengths Atacama Large Millimeter / submillimeter Array
(\ALMA) has recently resolved the cold molecular gas associated with
the outer regions of the torus in NGC~1068 and other nearby Seyferts
\citep{Garcia-Burillo+2016, Gallimore+2016, Imanishi+2016,
  Alonso-Herrero+2018, Alonso-Herrero+2019, Combes+2019}. In the
particular case of NGC~1068 ALMA observations found a torus no larger
than \about 11~pc in diameter with tracers of dense molecular gas
(e.g., $\rm HCO^{+}(4-3)$) and 2--3 times larger with lower-density
tracers such as $\rm CO(2-1)$ and $\rm CO(3-2)$
\citep{Garcia-Burillo+2019}, as well as evidence for nuclear molecular
outflow \citep{Gallimore+2016}. In addition, ALMA polarimetric
observations at 860 \mic\ with an angular resolution of $0 \farcs 07$
have measured the signature of magnetically aligned dust grains in the
torus of NGC~1068 \citep{Lopez-Rodriguez+2020}. These results
challenge the static nature of the torus in favor of an obscuring
structure generated by a hydromagnetic disk outflow, which has been
suggested from a theoretical point of view
\citep[e.g.,][]{Emmering+1992, Elitzur_Shlosman2006, Wada_2012,
  Elitzur_Netzer_2016, Dorodnitsyn+Kallman_2017}. Recently, more
detailed \VLTI\ measurements of two dozen nearby AGNs have complicated
the simple unification picture further. In several studied objects
most of the MIR emission appears to emanate from polar regions high
above the equatorial plane, i.e. not from where the bulk of the dust
is thought to reside \citep{Hoenig+2012, Hoenig+2013, Burtscher+2013,
  Lopez-Gonzaga+2016b, Leftley+2018}. \citet{Tristram+2014} suggested,
for the case of Circinus, that the elongation is due to seeing the
inner funnel wall of a torus slightly tilted toward us. Others propose
that the polar emission is caused by a dusty outflow and have
attempted to explain the observed elongations, e.g., by modifying the
distribution of dusty clouds in the models
\citep{Hoenig_Kishimoto_2017}, or through interplays between a tilted
accretion disk that anisotropically illuminates a hollow, biconical,
dusty wind \citep{Stalevski+2017, Stalevski+2019}.

In the near future extremely large single-aperture telescopes
(henceforth \emph{single-dish} for simplicity) such as the Giant Magellan
Telescope (\GMT), Thirty Meter Telescope (\TMT), and Extremely Large
Telescope (\ELT) will also be able to resolve the nuclear AGN thermal
emission in several nearby sources, as we will demonstrate in the
companion paper \citep[][hereafter Paper II]{hc-paper2}. \ALMA\
continues to improve its capabilities, yielding an evermore detailed
view of AGN tori at millimeter wavelengths \citep{Imanishi+2018,
  Garcia-Burillo+2019, Impellizzeri+2019}. Clearly, there is a strong
and growing demand for systematic studies and modeling of the resolved
AGN emission, i.e. imagery at all relevant wavelengths. Further
progress can arrive through brightness maps obtained from radiative
transfer in physically motivated dust distributions, but not through
models of the directly observed brightness distribution in the
sky. The latter have no connection to the astrophysical entities and
teach us little about the circumnuclear matter distribution.

In this paper we introduce in Section \ref{sec:hypercatsoftcubes} the
$n$-dimensional hypercubes of brightness and dust maps made with \C, a
radiative transfer code for clumpy media (\nenka, \nenkb). In
Section~\ref{sec:morphology}, using techniques based on image moments,
we investigate the morphologies of light and dust maps as a function
of model parameters and investigate how such models can generate
significant polar elongation in the MIR. We summarize the results in
Section \ref{sec:summary}.
A number of appendix sections and a User
Manual\footnote{\url{https://github.com/rnikutta/hypercat/blob/master/docs/manual/hypercat_user_manual.pdf}}
distributed with \HC\ shall serve as technical reference and guide
users in their work with the \HC\ software and model hypercubes.

In the companion \papertwo\ we will then apply the models to simulate
various synthetic observations of NGC~1068, including with 30~m class
telescopes.


\section{Hypercubes of Light Emission and Dust distribution}
\label{sec:hypercatsoftcubes}
\noindent
The need to model integrated-light SEDs from AGNs via radiative
transfer was recognized long ago. However, studies of resolved images
remain in their infancy. The reasons are twofold. First, until
recently no spatially resolved data were available. This has been
remedied by \VLTI\ and \ALMA, and new facilities such as \GMT, \TMT,
and \ELT\ will add new data in the near future. Second, the
computation, storage, and processing requirements for image sets with
parameter coverage equivalent to the model sets used in SED studies
are prohibitively expensive.

\HC\ is designed to break this second barrier. It is a user-friendly
software that abstracts away the underlying complexity of handling
very large $n$-dimensional hypercubes of data --- in this case images of
emission and projected dust distributions that are complex functions
of several independent model parameters. \HC\ enables one to generate
an image of the torus emission (or a dust distribution) for any
combination of parameters via multilinear interpolation in fractions
of a second on off-the-shelf standard computers. It can also simulate
observations of these images with single-dish telescopes, performing
convolution with any point-spread function (PSF; provided by the user
or computed by \HC), image transformations (e.g., rotation, scaling in
flux and size, resampling), noise addition to specified
signal-to-noise ratio (SNR), and image deconvolution. Interferometric
observations can also be simulated for a set of $(u,v)$ points,
generating synthetic visibility measurements, which will be presented
in a follow-up manuscript. Furthermore, spatially resolved maps of
spectral properties can be simulated, akin to integral field units
(IFUs). A module with various image morphology estimators completes the
set of high-level tools built into \HC. Descriptions of \HC\ software
interfaces, and of the (minimal) GUI program, are given in the User
Manual.

The software is open-source, written in \textsc{Python}, and designed
such that users can easily plug in their hypercubes of model images
and perform any analyses facilitated by \HC. We document in the User
Manual the exact requirements and provide a tool to create such
hypercubes from a collection of image files, and we can also assist
colleagues upon request.

With this first release of \HC\ we also release \emph{our} hypercube
of model images, generated with \C\ torus models. They constitute a
comprehensive set of thermal emission and dust distribution images,
which will allow to model and analyze data to be obtained by current
interferometric facilities and future 30-m class telescopes. We
describe the \C\ / \HC\ models in the remainder of this section.

\subsection{Distribution of Dust Clouds around the AGN}
\noindent
\C\ models the distribution of dust clouds around an AGN with an
axisymmetric function of cloud number density (per unit length) that
is separable in its radial and polar arguments $(r,\beta)$,
\begin{equation}
  \label{eq:Nc}
  \Nc(r,\beta) = C\, \frac{\NT(\beta)}{\Rd} \left(\frac{r}{\Rd}\right)^{\!-q},
\end{equation}
where $r$ is the radial distance from the AGN, $\beta$ is the angle of
a radial ray away from the equatorial plane, and $\NT(\beta)$ is the
mean number of clouds encountered along such a ray. The free parameter
\q\ sets how sharply \Nc\ falls off with distance.

The clouds, each with optical thickness \tv\ in the $V$ band, exist
between the dust sublimation radius
\begin{equation}
  \label{eq:Rd}
  \Rd\ = 0.4 \left(\dfrac{\Lb}{10^{45}\, {\rm erg\, s^{-1}}} \right)^{\!1/2} \left( \dfrac{T_{\rm sub}}{1500\, {\rm K}} \right)^{\!-2.6}\, {\rm (pc)},
\end{equation}
given in \nenkb\ and \citet{Barvainis1987}, and an outer radius
$R_{\rm out}$, with $\Y = R_{\rm out} / \Rd$ a free parameter. An
outer radius is not too meaningful a parameter, as it will arbitrarily
cut off the distant cloud distribution, but it is a necessary
parameter for the sake of numerical simulations.

\Rd\ is set by the AGN bolometric luminosity, \Lb, for a given dust
grain sublimation temperature, $T_{\rm sub}$ (1500~K for a standard
ISM \hbox{silicate/graphite} mix). The point source at the center
illuminates the dust isotropically with luminosity \Lb. $C$ in
Equation \eqref{eq:Nc} is a constant that ensures the correct normalization
of \hbox{$\NT(\beta) = \int\!\Nc(r,\beta)\, \dif r$}. The default
polar-angle density behavior in \C\ is a soft-edged Gaussian
%
\begin{equation}
  \label{eq:Nt}
  \NT(\beta) = \No \exp(-\beta^2/\sigma^2),  
\end{equation}
with \No\ the mean number of clouds per radial ray in the equatorial
plane and \sig\ the width parameter of the Gaussian, measured in
degrees from the equatorial plane. Other prescriptions for
$\NT(\beta)$ are possible. For instance, \citet{Fritz+2006},
\citet{Feltre+2012}, and \citet{Stalevski+2012} use a cosine in the
exponential. Whatever the exact nature of the polar edge of the torus,
observations require it to be soft and not sharp (\nenka).

All free parameters of the model geometry and the external viewing
angle of the observer form the vector of model parameters
$\bm{\theta} = (\sig, \iv, \Y, \No, \q, \tv)$. Their sampling in \HC\
is listed in Table~\ref{tab:parameter_sampling}.

\begin{deluxetable}{llclr}
  \tablecaption{Parameter Sampling\label{tab:parameter_sampling}}
  \tablewidth{0pt}
  \tablehead{
    \colhead{Param.} & \colhead{Sampled Values} & \colhead{Units} & \colhead{Dep.$^a$} & \colhead{$N_{\theta_k}$$^b$}
  }
  \startdata
  \sig\          & 15, 30, 45, 60, 75           & deg   & b, d  & 5  \\[-2pt]
  \iv\           & 0, 10, $\ldots$ , 90         & deg   & b, d  & 10 \\[-2pt]
  \Y\            & 5, 6, $\ldots$ , 20          &       & b, d  & 16 \\[-2pt]
  \No\           & 1, 2, $\ldots$ , 12          &       & b$^c$ & 12 \\[-2pt]
  \q\            & 0, 0.5, 1, 1.5, 2            &       & b, d  & 5  \\[-2pt]
  \tv\           & 10, 20, 40, 60, 80, 120, 160 &       & b     & 7  \\[-2pt]
  wave$^d$       & 1.2, $\ldots$ , 945          & \mic  & b     & 25 \\
  \hline
  \multicolumn{4}{l}{Total number of combinations without wave:} & 336 k \\
  \multicolumn{4}{l}{\phantom{Total number of combinations} with wave:}    & 8.4 M \\
  \enddata
  \tablecomments{$^a$Maps that depend on this parameter: b =
    brightness map, d = dust map.
    $^b$Number of sampled values for each parameter.
    $^c$\No\ is only a multiplicative factor (see text).
    $^d$See Table~\ref{tab:wavelengths} for wavelength sampling. Other
    parameter values, for instance, larger \Y, can be added in the
    future based on community needs.}
\end{deluxetable}

Note that Equation \eqref{eq:Nc} prescribes a \emph{continuous} cloud
number density at any location $(x,y,z)$. Unlike any other published
model of clumpy tori, \C\ does not place macroscopic clouds at
discrete locations within the computational domain, i.e. it does not
introduce clumpiness via Monte Carlo simulations. Instead, clumpiness
is dealt with analytically by computing the local statistical
properties of the stochastic medium. This is computationally very
fast but produces images of smooth appearance, despite the clumpiness
of the dust. The computed images are simply the limit (mean) of
infinitely many random realizations of a discrete model.

We account for two classes of clouds: those irradiated \emph{directly}
by the central source (their source function is
$S_{\!\lambda}^{\rm dir}$), and those that do not have a clear LOS to
the AGN and are heated only \emph{indirectly} by an isotropic
radiation field provided by other clouds around it (their source
function is $S_{\!\lambda}^{\rm ind}$). The directly illuminated
clouds have a hot, AGN-facing side and a cooler, averted side. Their
source function is therefore highly anisotropic. An observer will see
a mix of both the hot and cool sides, depending on the position angle
with respect to a cloud, like the phases of the Moon. We precompute
the cloud emission source function \Sc\ for a finely sampled range of
distances $r$ from the AGN (to account for heating by the incident
flux at that distance) and position angles $\alpha$ (to account for
the anisotropic illumination). At all other locations the source
function is smoothly interpolated. The overall cloud source function
(see Equation 10 in \nenka) at any location is a mix of the two cloud types
\begin{equation}
  \label{eq:sfn_overall}
  S_{C,\lambda}(r,\alpha,\beta) = p(r,\beta)\, S_{\!\lambda}^{\rm dir}(r,\alpha) + [1-p(r,\beta)]\, S_{\!\lambda}^{\rm ind}(r)
\end{equation}
with the mixing proportion being set by the probability $p(r,\beta)$
that a cloud at location $(r,\beta)$ has a clear view of the AGN. This
probability is $\exp\{-\N(r,\beta)\}$, with
$\N(r,\beta) = \!\int_0^r \Nc(r,\beta)\, \dif r$ the mean number of
clouds between the cloud and the AGN. The observed brightness map,
integrated flux, and SED are then computed a posteriori and very
quickly via ray-tracing through a distribution of clouds. We describe
this in the next subsection.

\subsection{Monochromatic Emission Maps}
\label{sec:brightnessmaps}
\begin{figure*}
  \includegraphics[width=0.445\hsize]{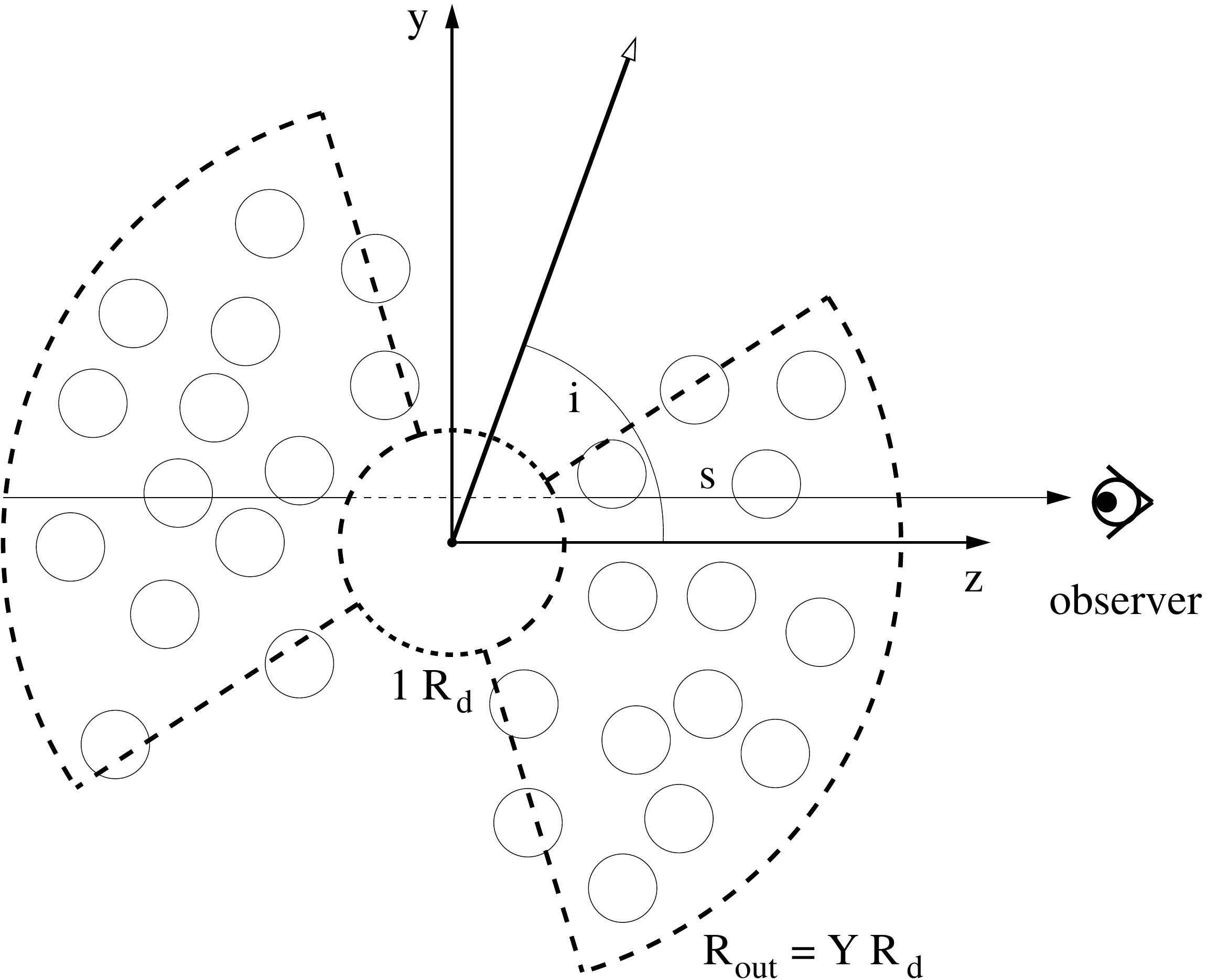} 
  \includegraphics[width=0.54\hsize]{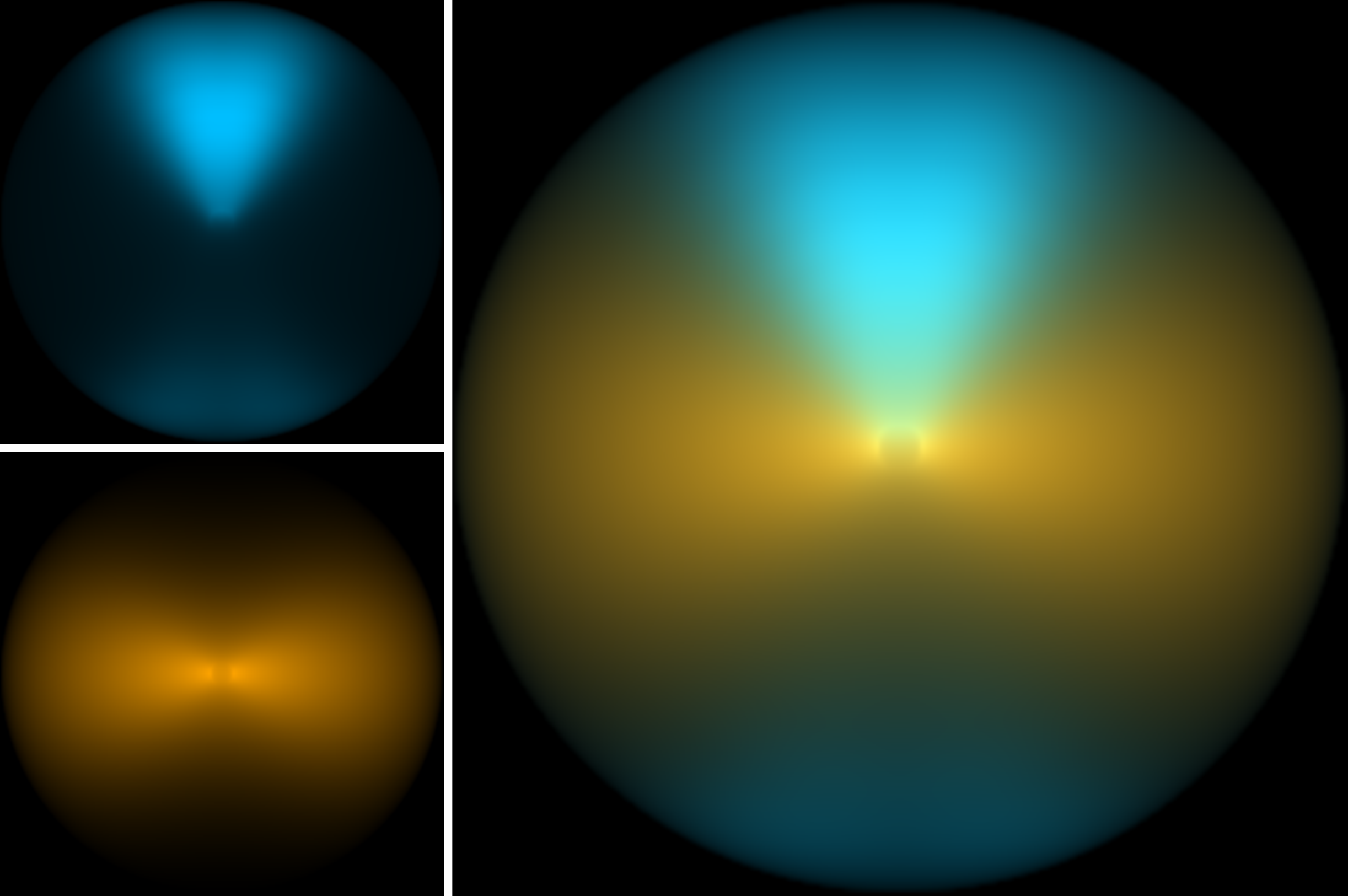} 
  \caption{Left: vertical cut through a schematic toroidal
    cloud distribution, illustrating integration along a particular
    path $s$ parallel to the $z$-axis. Dust-free regions such as the
    central cavity of $1\,\Rd$ diameter do not contribute to the
    integral. The torus angular edge is soft, i.e. the cloud number
    falls off from the midplane like a Gaussian. Note that \C\ does
    not realize discrete clouds, but rather the properties of a clumpy
    medium at any continuous location. Right: what an observer
    on the left might see; two \C\ thermal emission maps at 9.5~\mic\
    (MIR, top panel, blue colors) and 860~\mic\ (submillimeter, bottom
    panel, orange colors), and their composite (large panel),
    generated by the same cloud distribution. The model has parameters
    $\sig=43$\degr, $\iv=75$\degr, $\Y = 20$, $\No = 10$, $\q = 0.08$,
    and $\tv = 70$. The relative contributions of the two maps to the
    composite image were normalized for clarity.}
  \label{fig:zintegration}
\end{figure*}
\noindent
Figure~\ref{fig:zintegration} shows the geometry of ray-tracing through
a \C\ torus model. The plane of the sky is $(x,y)$, with $y$ pointing
up and $x$ perpendicular to the page (pointing away from the
reader). The AGN-to-observer LOS is $z$. The torus axis is
in the $(y,z)$-plane, inclined from the \zax\ by an angle \iv. At
every coordinate $(x,y)$ in the plane of the sky \C\ computes the
monochromatic intensity $I_\lambda(x,y)$ along the path $s$ (parallel
to the \zax) as the integral over a product of three functions
\begin{equation}
  \label{eq:I}
  I_\lambda(s;x,y) = \int\!\dif s\, \Sc\, \Nc(s')\, P_{\rm esc}(s',s).
\end{equation}
$\Sc(s')$ is the photon-generating cloud source function at a given
point $s'$ along the path. $\Nc(s')$ is the local cloud number density
from Equation \eqref{eq:Nc} and describes how many clouds are generating
the photons. Finally,
\begin{equation}
  \label{eq:Pesc}
  P_{\rm esc}(s',s) = \exp{\left\{- {\cal N}(s',s)(1-\exp{(-\tau_\lambda)})\right\}}
\end{equation}
is the probability that a photon generated at $s'$ propagates all the
way along the rest of the path to the observer without being absorbed
again, assuming that the number of clouds is Poisson distributed
around the mean \citep{NattaPanagia1984}.
\hbox{${\cal N}(s',s) = \int_{s'}^s \Nc\, \dif s$} is the mean number
of clouds between point $s'$ and the observer, and $\tau_\lambda$ the
optical depth of a single cloud at wavelength $\lambda$. The
convergence of Equation \eqref{eq:I} is ensured by \C\ through Romberg
integration.

Ray-tracing for all sky positions $(x,y)$ in the field of view (FOV)
generates a 2D image $I_\lambda(x,y)$ at wavelength
$\lambda$. Figure \ref{fig:zintegration} shows two such maps, and their
composite, at 9.5~\mic\ (blue) and 860~\mic\ (gold).
Correct normalization of these brightness maps, as described
in Appendix A of \nenkb, requires convergence of the area integral
$\int\!\dif x\, \dif y\,I_\lambda(x,y)$; \C\ handles this via the
``Cuhre'' integration with globally adaptive subdivision, provided by
the \textsc{Cuba} library \citep{HAHN200578}.

\subsection{Wavelength Axis and SEDs}
\label{sec:waveaxis}
\noindent
\C\ computes thermal emission maps $I_\lambda(x,y)$ at all specified
wavelengths independently. Their area integrals
(Section \ref{sec:brightnessmaps}) and proper normalization produce the
SEDs that have been available since 2008 (and are being updated from
time to time) for a large number of model parameter
combinations.\footnote{\C\ torus SEDs can be found at
  \href{https://www.clumpy.org}{www.clumpy.org}}
\begin{deluxetable}{cccccrcl}
  \tablecaption{Images Computed at the Following Wavelengths.\label{tab:wavelengths}}
  \tablewidth{0pt}
  \tablehead{
    \colhead{No.} & \colhead{Wave}   & \colhead{Band} & \colhead{\phantom{a}} & \colhead{No.} & \colhead{Wave}   & \colhead{Band} & \colhead{}\\
    \colhead{}    & \colhead{(\mic)} & \colhead{}     & \colhead{}            & \colhead{}    & \colhead{(\mic)} & \colhead{}     & \colhead{}
  }
  \decimals
  \startdata
  0       &  1.2 & J &  & 13  &  12.5\phantom{$^a$} & N            &                     \\[-2pt]
  1       &  2.2 & K &  & 14  &  18.5\phantom{$^a$} & Q            &                     \\[-2pt]
  2       &  3.5 & L &  & 15  &  31.5\phantom{$^a$} & Z            &                     \\[-2pt]
  3       &  4.8 & M &  & 16  &  37.1\phantom{$^a$} & Z            &                     \\[-2pt]
  4       &  8.7 & N &  & 17  &  53\phantom{$^a$}   & A            &                     \\[-2pt]
  5       &  9.3 & N &  & 18  &  89\phantom{$^a$}   & C            &                     \\[-2pt]
  6       &  9.8 & N &  & 19  & 154\phantom{$^a$}   & D            &                     \\[-2pt]
  7       & 10.0 & N &  & 20  & 214\phantom{$^a$}   & E            &                     \\[-2pt]
  8       & 10.3 & N &  & 21  & 350$^a$             & 10           & (857 GHz)$^b$ \\[-2pt]
  9       & 10.6 & N &  & 22  & 460$^a$             & \phantom{1}9 & (652 GHz)$^b$ \\[-2pt]
  10      & 11.3 & N &  & 23  & 690$^a$             & \phantom{1}8 & (434 GHz)$^b$ \\[-2pt]
  11      & 11.6 & N &  & 24  & 945$^a$             & \phantom{1}7 & (317 GHz)$^b$ \\[-2pt]
  12      & 12.0 & N &  &     & \phantom{}          &              &                     \\
  \enddata
  \tablecomments{ $^a$At center of bandpass, computed in wavelength
    space. $^b$\ALMA\ bandpass central frequency. See Cycle-6
    Handbook:
    \url{https://almascience.nrao.edu/documents-and-tools/cycle6/alma-technical-handbook}}
\end{deluxetable}

For this release of \HC\ we have sampled \hbox{$N_\lambda = 25$}
wavelengths, listed in Table~\ref{tab:wavelengths}. The wavelengths
were chosen strategically based on these criteria: (i) cover all the
important bands by available and upcoming instruments, (ii) ensure a
spectrally well-sampled 10~\mic\ silicate feature, (iii) enable
interpolation between wavelengths, including in the Raleigh-Jeans tail
of the SEDs. However, we needed to limit the number of offered
wavelengths to keep the data size of the hypercube manageable. Note
that the images are monochromatic, which is justified under the
assumption of very narrowband filters commonly used in IR observations
of AGNs.\footnote{See, e.g., the filters defined for \emph{CanariCam}
  (\url{http://www.gtc.iac.es/instruments/canaricam/\#Filters}) or
  \emph{T-ReCS}
  (\url{http://www.gemini.edu/sciops/instruments/trecs/imaging/filters})}.

We point out that although \HC\ has a discrete sampling of
wavelengths, images at any wavelength within the 1.2--945~\mic\ range
can be computed through interpolation. An approximate SED can be
easily extracted, interpolating if necessary between the 25 sampled
wavelengths in the hypercube (see User Manual). For studies geared
toward SEDs that require a higher spectral resolution, we refer the
reader to the established rich grid of model SEDs available on the \C\
website. These model SEDs have been successfully exploited since their
introduction in \nenka\ and \nenkb, in most cases to fit the observed
nuclear but unresolved emission of AGNs
\citep[e.g.,][]{AsensioRamos2009, Nikutta+2009, Alonso-Herrero+2011,
  RamosAlmeida+2011, Privon+2012, Ichikawa+2015, Sales+2015, Fan+2016,
  Fuller+2016, Xie+2016, Audibert+2017, Mateos+2017, Ricci+2017,
  Garcia-Bernete+2019, Gonzalez-Martin+2019a, Gonzalez-Martin+2019b}.

\subsection{Simplifications and Caveats}
\label{sec:caveats}
\noindent
The \C\ models, as well as the underlying \D\ radiative transfer models,
make certain simplifying physical assumptions. The present work does
not aim to modify these computations; thus, \HC\ inherits their
limitations. Here we note the most important of these.

As published, the \C\ models represent a single set of dust grain
optical properties -- a standard ISM mix of cold astrophysical
silicates \citep{OHM1992} and graphites \citep{Draine2003} -- and a
parameterized power-law grain size distribution following
\cite{MRN_1977}. The ``cold'' interstellar silicates from
\cite{OHM1992} are especially successful in fitting the silicate
depths observed in AGNs, at both 10 and 18~\mic\ \citep{Nenkova+2008a,
  Sirocky+2008, Thompson+2009}, but we note that changing the optical
and physical properties can result in significantly different SED
shapes.

\C\ models can underpredict the amount of near-IR (NIR) emission in
type~1 sources \citep{Mor+2009} because the averaged composite grains
have a dust sublimation temperature of around 1500~K, whereas larger
and more graphite-rich grains can survive up to $\sim$1800~K, thus
producing more NIR emission.

\C\ models do not produce strong silicate emission features and never
show very deep silicate absorption. In the context of AGN tori this is
a \emph{feature} of clumpy torus models. Observations of
\emph{nuclear} emission never detected deep silicate absorption. All
sources that do exhibit deep silicate absorption either are ULIRGs
\citep{Levenson+2007} or suffer LOS absorption along their (host)
galactic plane \citep{Goulding+2012}.

\C\ models are computed with a statistical (Poisson) distribution of
dusty clouds around a mean value (the parameter \No), and the resulting
radiative transfer solutions are asymptotic averages over many
discrete realizations of such distributions. But \C\ models are
\emph{not} single-instance discrete realizations, unlike, e.g., Monte
Carlo models. The benefit is extremely fast model computation. The
drawbacks are that model images appear smooth despite being computed
for a clumpy medium and that no stochastic deviations from the mean
solution can be studied with such models.

The computation of \C\ cloud source functions involves an iterative
scheme between the clouds being directly heated by the AGN and clouds
whose LOS to the AGN is blocked and that are only heated by the
radiation bath provided by other clouds in their vicinity. \C\
executes only the first two steps of this iteration. This
approximation is valid for volume filling factors $<0.1$
\citep{Nenkova+2008a}, which, depending somewhat also on other
parameters, holds for $\No\lesssim12$.

The largest tori in the images released with \HC\ have a radial extent
of $\Y=20$ dust sublimation radii. If these were used to compare with
larger tori in nature, the amount of far-IR (FIR) emission from heated
dust might be underestimated. It is important to note that at FIR and
submillimeter wavelengths there are often significant contributions
from other processes, e.g., star formation and synchrotron
radiation. Thus, modeling should include these components
explicitly. In this work we concentrate on the possibility of resolved
emission at MIR wavelengths, that is, tori with $\Y\leq20$ are
sufficient.

\subsection{Image Size and Sampling}
\label{sec:imagesize}
\noindent
Since the 3D cloud distribution is azimuthally symmetric, the 2D
brightness distribution in the $(x,y)$-plane is always symmetric about
the \yax\ (left-right symmetry), regardless of inclination $i$.
\C\ exploits this left-right symmetry when computing a brightness map
and only ray-traces one half of the image.
If requested, it can mirror the computed image about the \yax\ and
output the full $N_x \times N_y$ pixel image.

In a full-size (square) image, $N_x = N_y$ and always \emph{odd}, to
ensure that the nuclear unresolved source is chiefly located at the
single central pixel.
The central source is unresolved under all circumstances.
With \etapix\ the user-requested image resolution in units of pixels
per \Rd, a full-size image of radial extent \Y\ (in units of \Rd) has
dimensions $N_x = N_y = 2\, Y \etapix + 1$ pixels.
A half-size image has dimensions \hbox{$N_x = Y \etapix + 1$} and
\hbox{$N_y = 2\, Y \etapix + 1$} pixels.

Note that the image size (in pixels) depends on one of the parameters
of the model: the torus radial extent \Y. That is, by default the
torus image computed by \C\ always fills the image frame.
This is desirable for modeling work but may not be ideal when
simulating observations where the FOV is usually much larger
than the small AGN torus on the sky.
A constant image size (in pixels) irrespective of \Y\ is also
necessary for creating from all single images an $n$-dimensional
hypercube. We therefore instructed \C\ to embed all tori within an
image frame of constant FOV, regardless of their size. This common
size is the largest sampled \Y\ value in our hypercube,
$\Ymax=20$. \C\ writes the emission maps to \texttt{FITS} files (one
file per model parameter combination, one 2D slice per wavelength).

\subsection{Projected Cloud Number Maps}
\label{sec:cloudmap}
\noindent
While ray-tracing, \C\ also computes the integrated number of clouds
along the LOS. For every path $s$ parallel to the \zax\ in the FOV
(i.e. every pixel) this number is ${\cal N}(s) = \int\! \Nc\, \dif s$,
with \Nc\ from Equation \eqref{eq:Nc}. All pixels together make up a map of
the projected dust cloud number distribution, $C_{d}(x,y)$. Because of
the same argument about azimuthal symmetry made in
Section \ref{sec:imagesize}, $C_{\rm d}$ is always left-right symmetric
in the image plane, regardless of inclination $i$.

Unlike the emission maps $I_\lambda(x,y)$ the cloud maps $C_{d}$ do
not depend on $\lambda$ or \tv. Furthermore, from Eqns.~\eqref{eq:Nc}
and \eqref{eq:Nt} we see that the cloud distributions at fixed
$\left\{\sig, \iv, \Y, \q\right\}$ are identical in structure for all
\No\ up to a factor: \No\ itself. We can thus store just one cloud map
$C_{d}(\No\!\!=\!\!1)$ and later multiply it by the desired \No\ when
using the cloud maps. This greatly reduces the storage requirements
for all cloud maps.

\begin{figure*}
  \includegraphics[width=\hsize]{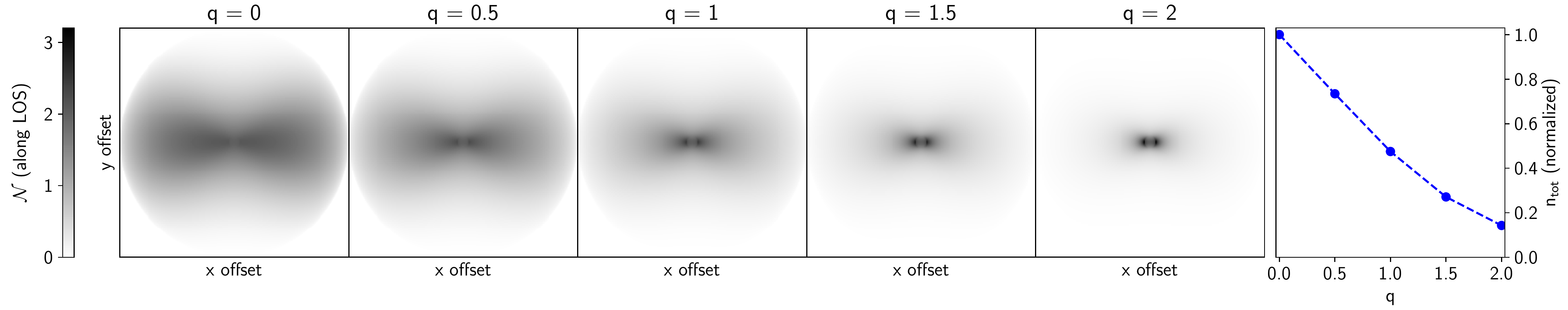}
  \caption{Projected maps of the number of clouds \N\ along
    LOSs for a model with parameters within \q\ = 0--2 as
    indicated, and other parameters held fixed at \sig\ = 43\degr,
    \iv\ = 90\degr, \Y\ = 20, and \No\ = 1. All maps are normalized to
    the same maximum $\N_{\rm max} = 3.2$ (which occurs at \q\ = 2)
    and they share the gray scale on the left. The right panel shows
    the total number of clouds $n_{\rm tot}$ in each map as a function
    of \q, normalized to the peak (at \q\ = 0).}
  \label{fig:Ntot_vs_q}
\end{figure*}

The integral over all pixels in the cloud map yields the total number
of clouds in the torus,
\hbox{$n_{\rm tot} = \int\! C_{d}(x,y)\, \dif x\, \dif y$}. This
number is preserved no matter what viewing angle the torus is inclined
at. Note that $n_{\rm tot}$ changes with \q. From Section 2.4 of
\nenka, $n_{\rm tot} = (\Nc/A_C)\, \dif V$, where the cloud cross
section $A_C \simeq R_C^2$ (and $R_C$ is the cloud size).  From
Equation \eqref{eq:Nc}, $\Nc \propto r^{-q}$. With fixed \Rd, $R_C$ and
\No\ we then obtain $n_{\rm tot} \propto r^{-q}\, \dif r$, i.e. when
\q\ increases, $n_{\rm tot}$ decreases. Figure~\ref{fig:Ntot_vs_q}
shows in the first five panels the projected cloud numbers per pixel
(per LOS) for a torus with five values of \q\ between 0 and 2, with
other parameters fixed at \sig\ = 43\degr, \Y\ = 20, \No\ = 1, and at
a viewing angle \iv\ = 90\degr. The radial cloud distribution is
visibly concentrated with growing \q. At the central pixel, i.e. at
zero $x$ and $y$ offsets from the center, \N\ = 2 always, because \No\
= 1, i.e. there is on average one cloud along a radial ray in the
equatorial plane (and one on the far side of the torus). The right
panel shows $n_{\rm tot}$ as a function of \q\ normalized to the maximum
value of $n_{\rm tot}$, which occurs at \hbox{\q\ = 0}.

These physically motivated maps of dust distributions, when compared
to the morphology of the thermal emission emanating from the same
torus, can help to understand the conditions that cause the polar
elongation of the MIR emission observed in some AGN (or equatorial
elongations in other sources, for that matter). The behavior of
morphology as a function of position angle (\pa), wavelength, and all
model parameters can be studied. In Section~\ref{sec:morphology} we
analyze the dust and emission morphologies in detail.

\subsection{Hypercube Model Files}
\label{sec:compstorage}

\noindent
The parameter sampling of our \C\ hypercubes is listed in
Table~\ref{tab:parameter_sampling}. The total number of model
realizations, without wavelength, is
$\prod_k\! N_{\theta_k} = 336,\!000$. The images were computed for
$N_\lambda = 25$ wavelengths, listed in Table~\ref{tab:wavelengths}
(see also Section~\ref{sec:waveaxis}); thus, 8.4 million images were
computed overall, in addition to 4000 individual dust
maps. Computation of the \C\ brightness maps and dust cloud maps that
make up this first release of the hypercube consumed a total of
3.2 yr on mid-tier CPUs.

We deliver all computed brightness maps stored as a 9D hypercube
within convenient \texttt{hdf5} files, in the \texttt{imgdata}
group. The same files also contain a 6D hypercube of all projected
dust maps (\texttt{clddata} group). See Table~\ref{table:hdf5files}
and Appendix~\ref{app:hdf5organization} for details about the
\texttt{hdf5} files and their internal organization.

\begin{deluxetable*}{lrrrl}
  \tablecaption{List of \texttt{hdf5} Files Containing \C\ Brightness Maps and Projected Dust Maps.\label{table:hdf5files}}
  \tablewidth{0pt}
  \tablehead{
    \colhead{File Name} & \multicolumn2c{Size (GB)}                        & \colhead{Nwave} & \colhead{Wavelengths$^{a}$} \\
    \colhead{}          & \colhead{Compressed$^{b}$} & \colhead{Raw$^{c}$} & \colhead{}      & \colhead{(\mic)}
  }
  \startdata
  \texttt{hypercat\_20200830\_all.hdf5}   & 271\phantom{00} & 913\phantom{00} & 25 & all of the below \\
  \texttt{hypercat\_20200830\_nir.hdf5}   &  44\phantom{00} & 146\phantom{00} &  4 & 1.2, 2.2, 3.5, 4.8 \\
  \texttt{hypercat\_20200830\_mir.hdf5}   & 120\phantom{00} & 402\phantom{00} & 11 & 8.7, 9.3, 9.8, 10, 10.3, 10.6, 11.3, 11.6, 12, 12.5, 18.5 \\
  \texttt{hypercat\_20200830\_fir.hdf5}   &  65\phantom{00} & 219\phantom{00} &  6 & 31.5, 37.1, 53, 89, 154, 214 \\
  \texttt{hypercat\_20200830\_submm.hdf5} &  42\phantom{00} & 146\phantom{00} &  4 & 350, 460, 690, 945 \\
  \enddata
  \tablecomments{$^{a}$Note that \HC\ can interpolate between the wavelengths. $^{b}$To be downloaded. $^{c}$Uncompressed on disk. Files available at \url{https://www.clumpy.org/images/}}
\end{deluxetable*}
%


\section{Morphologies of Thermal Emission and Projected Cloud Number Maps}
\label{sec:morphology}
\noindent
The emergence of instruments powerful enough to resolve the dusty
torus in nearby AGNs (e.g., \ALMA, 30~m class telescopes) allows us not
only to separate the flux contributions of the nucleus and host galaxy
but also to quantify the morphology of the observed nuclear brightness
distribution. Resolved \VLTI\ observations \citep[e.g.,][]{Hoenig+2012,
  Hoenig+2013, Tristram+2014, Lopez-Gonzaga+2016b} show that in
several nearby AGNs the MIR emission is elongated in polar directions,
with the polar axis usually defined by other means, e.g., the
orientation of ionization cones, a nuclear jet, or the perpendicular
orientation of a maser disk. In these sources most of the MIR emission
appears to be emanating from regions high above the equatorial plane
of the ``torus,'' i.e. not from where most of the dust is thought to
reside according to AGN unification.

Whatever the mechanism for producing an elongated morphology, it must
reconcile the 2D brightness map with the 3D dust distribution
consistently. NIR and MIR imaging with extremely large single-dish
telescopes will resolve the morphology unambiguously, and free of
model assumptions. Phenomenological modeling of the brightness
distribution can no longer serve as a surrogate.

In this section we investigate the morphologies of \C\ brightness maps
in several wavelength regimes and of their underlying dust
distribution maps. We introduce and measure various morphological
quantities, e.g., centroid location, half-light radii (1D), radii of
gyration (2D), elongation (aspect ratio) of the morphology, position
angle, asymmetry (skewness), and compactness (peakedness). We first
investigate the morphological changes to brightness maps as a function
of model parameters, with a focus on identifying the part of parameter
space that allows for polar elongation of brightness distributions
similar to those reported in the literature. We then compare these
brightness maps to their corresponding dust maps, inferring as a
function of model parameters the localized correspondence (or lack of
it) of dust distributions and the emission morphologies generated by
them. In \papertwo\ we will then simulate observations with extremely
large telescopes and compare to observational results obtained for
NGC~1068.

\subsection{Image Moments and Moment Invariants}
\label{sec:image-moments}
\noindent
\begin{deluxetable}{l}
  \tablecaption{Useful Morphological Measurements with Moments.\label{table:usefulmoments}}
  \tablehead{\colhead{Measure / Method}}
  \startdata
  Image mass (sum), or ``flux''\\[1pt]
     \qquad $M_{00}$ or $\mu_{00}$ (identical) \\[6pt]
  Image centroid location\\[1pt]
     \qquad $\bar x = M_{10}/M_{00}$,\, $\bar y = M_{01}/M_{00}$ \\[6pt]
  Radius of gyration about axis\\[1pt]
     \qquad $\Rgx = \sqrt{\mu_{20}/\mu_{00}}$,\, $\Rgy = \sqrt{\mu_{02}/\mu_{00}}$ \\[6pt]
  Elongation in the $y$-direction\\[1pt]
     \qquad $e=\Rgy\, /\, \Rgx$ \\[6pt]
  Image covariance matrix\\[1pt]
     \qquad $\mathrm{Cov}[I(x,y)]=\begin{bmatrix}\mu'_{20} & \mu'_{11}\\ \mu'_{11} & \mu'_{02}\end{bmatrix}$\\[6pt]
  \pa\ of main component\\[1pt]
     \qquad $\dfrac{1}{2}\arctan\left\{2\,\mu'_{11} \Big/ \left(\mu'_{20}-\mu'_{02}\right)\right\}\ \rm(radian)$\\[6pt]
  Skewness\\[1pt]
     \qquad $S_x=\mu_{03}\Big/\sqrt{\mu_{02}^3}$,\, $S_y=\mu_{30}\Big/\sqrt{\mu_{20}^3}$\\
  \enddata \tablecomments{ $M_{\rm pq}$ raw moment
    (Equation \ref{eq:appendix-momentraw}), $\mu_{\rm pq}$ central moment
    (Equation \ref{eq:momentcentral}),
    \hbox{$\mu'_{\rm pq} = \mu_{\rm pq} / \mu_{00}$} second-order central
    moment (see Appendix~\ref{sec:appendix-covariance}), $I(x,y)$
    image.}
\end{deluxetable}

\noindent
We quantify the morphologies in terms of their image moments and
quantities computed from moments, and we compare them to some traditional
morphological quantifiers. The so-called \emph{central moments} are
\begin{equation}
  \label{eq:momentcentral}
  \mu_{\rm pq} = \sum_x \sum_y I(x,y)\, (x - \bar x)^p\, (y - \bar y)^q, \quad p,q \in \mathbb{N},
\end{equation}
where $\bar x$ and $\bar y$ are the image centroid coordinates in
pixels. By construction central moments are invariant under
translation operations. Combinations of moments can be defined that
are invariant under translation, scaling, rotation, or any combination
of these operations. We refer to
Appendix~\ref{sec:appendix-morphology} for a technical
discussion. Functions of image moments can be used to compute various
quantities that characterize morphology. Table
\ref{table:usefulmoments} lists the ones that we make use of in this
paper.

\subsection{Size, Elongation, and Image Compactness}
\label{sec:size-elongation}
\noindent
We wish to quantify the spatial extension of emission morphologies as
a function of model parameters. Various measures have been proposed in
the literature, e.g., the half-light radius, the FWHM of a fitted 2D
Gaussian, or the Petrosian radius \citep{Petrosian_1976,
  Graham+2005}. We show that a quantity based on image moments, the
\emph{radius of gyration}, is ideally suited to measure morphology
size. We also introduce the \emph{Gini index} as a simple way to
quantify image compactness. For the remainder of the paper we can
assume that there is only one extended source present in the image --
the model torus.

\subsubsection{Half-light Radius}
\label{sec:half-light-radius}
\begin{figure*}
  \center
  \includegraphics[width=\hsize]{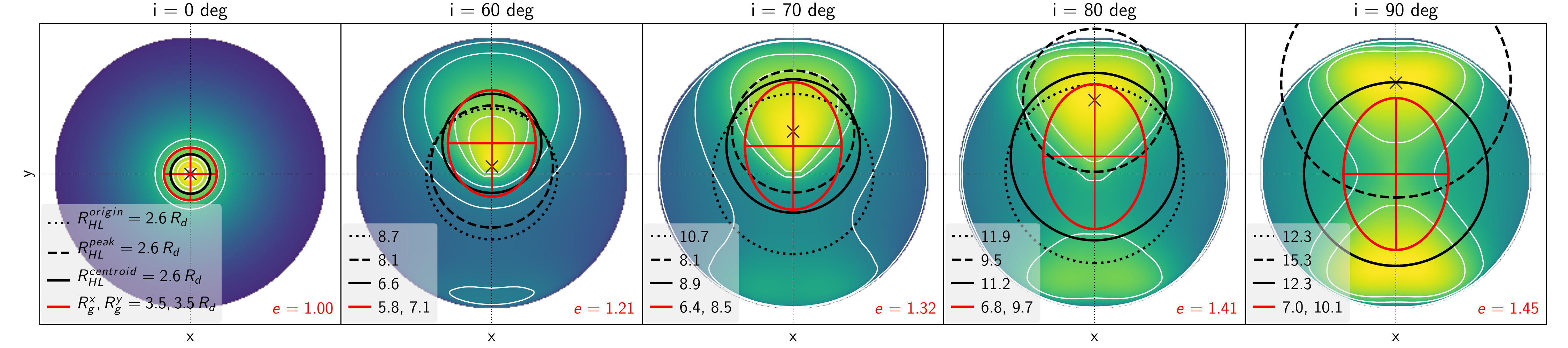}
  \caption{Morphological size measurements on 10~\mic\ images of a
    model at several inclination angles \iv\ (columns). Inclinations
    between 0\degr\ and 60\degr\ are similar in appearance and for
    clarity not shown. \No\ is set at 9, and all other parameters are
    fixed at the best values from SED fitting of NGC~1068 given in
    \citet{Lopez-Rodriguez+2018}: \sig = 45\degr, \Y = 18, \q = 0.08,
    \tv = 70, \iv = 75\degr. These values are also used throughout
    \papertwo. The $Y = 18~\Rd$ torus is embedded in a
    $2 \times 20~\Rd$ FOV frame. Images are log-stretched and
    normalized independently. White contour lines computed on the
    linearly scaled images are shown at 5\%, 10\%, 30\%, 50\% of the
    peak pixel, which is marked with a black cross. Red horizontal and
    vertical lines show the radii of gyration \Rgx, \Rgy\ about the
    $x$ and $y$-axes; they form the semimajor axes of the red ellipse
    encompassing them, centered on the image centroid
    $(\bar x, \bar y)$. For pole-on view (\iv\ $= 0$\degr) we set the
    peak location to the image origin. Three black circles show
    apertures with half-light radii centered on the origin (dotted),
    emission peak (dashed), and the centroid (solid). All three
    contain half the light in an image. In pole-on view all three
    circles are centered on the origin and thus of equal radius. In
    edge-on view it depends on the parameter combination whether the
    peak pixel will be at the origin or higher above the equatorial
    plane (as is the case here, with two symmetric peak pixels). The
    legend lists the half-light and gyration radii in units of
    \Rd. The elongation $e = \Rgy/\Rgx$ is printed in the lower right
    corner.}
  \label{fig:sizes}
\end{figure*}

\noindent
A commonly used size estimate of an emission region is the half-light
radius \RH\ of a circular aperture that contains half of the total
flux in the image (see
Appendix~\ref{sec:appendix-halflightradius}). The aperture is usually
anchored on the origin of the coordinate frame. However, the
half-light radius is sensitive to where the aperture is centered in
the general case that the brightness distribution is not
uniform. Figure \ref{fig:sizes} shows a 10.0~\mic\ image of a \C\ torus
model at several viewing angles \iv, as indicated. All other model
parameters are fixed.
Three circular apertures are shown, centered at different loci, and
each contains half the light of the image. Because the brightness
distribution is not uniform across the image, these half-light radii
differ in size.

In pole-on view (\iv\ = 0\degr) all half-light radii are identical,
since all apertures are centered on the origin. Note that in pole-on
view the ``brightest pixel'' is ill-defined; it is in fact a ring
around the image origin. We thus set the aperture center to the center
pixel. In edge-on views the three circles are only identical if the
brightest pixel happens to be at the image center. This is the case if
the central LOS is optically thin or mildly obscured. If the
absorption along the central LOS increases, through either a higher
\No\ or a higher \tv, or through a change in wavelength, the emission
peak occurs spatially much higher above the equatorial plane, and thus
the half-light aperture constructed around the peak pixel is very
different from the centroid- and origin-based ones.

Unresolved measurements of the total flux are of course not affected
by the choice of the aperture center, since all flux is within the
PSF. Interferometers without phase closure, such as \VLTI\ with two
telescopes, can resolve the nuclear emission in some nearby AGNs, but
the emission cannot be localized within the FOV since absolute
astrometry cannot be achieved. However, interferometers with phase
closure (e.g., \VLTI\ with MATISSE or GRAVITY, \ALMA) and, of course,
very large single-aperture telescopes (e.g., \GMT, \TMT, \ELT) do
attain absolute astrometry; thus, the locus at which the half-light
radius aperture is anchored can be important. We note though that for
extragalactic observations of (marginally) resolved nuclear cores,
precise image registration is critical \citep{Prieto+2014}.

In the edge-on scenario shown in Figure \ref{fig:sizes}, the offset
between the image origin and the brightest pixel is about 12.3
\Rd. The physical size depends on luminosity; for NGC~1068 it is quite
uncertain. Our own fitting (in \papertwo) of the $K$-band image by
\citet[][hereafter G20]{gravity2020} yields
${\Lb = 1.56 \times 10^{44}\, \rm erg\,s^{-1}}$ or
${\Lb = 2.62 \times 10^{44}\, \rm erg\,s^{-1}}$ depending on the
model; \cite{Garcia-Burillo+2014} find
${\Lb = 4.2 \times 10^{44}\, \rm erg\,s^{-1}}$ from SED fitting;
\cite{Lopez-Rodriguez+2018} report
${\Lb = 5.02 \times 10^{44}\, \rm erg\,s^{-1}}$ when including FIR
data (see also \papertwo); \cite{Alonso-Herrero+2011} adopted
${\Lb = 1 \times 10^{45}\, \rm erg\,s^{-1}}$; \cite{Raban+2009} give
${\Lb = 1.5 \times 10^{45}\, \rm erg\,s^{-1}}$; \cite{Burtscher+2013}
list ${\Lb = 1.58 \times 10^{45}\, \rm erg\,s^{-1}}$.

\begin{figure}
  \center
  \includegraphics[width=1\hsize]{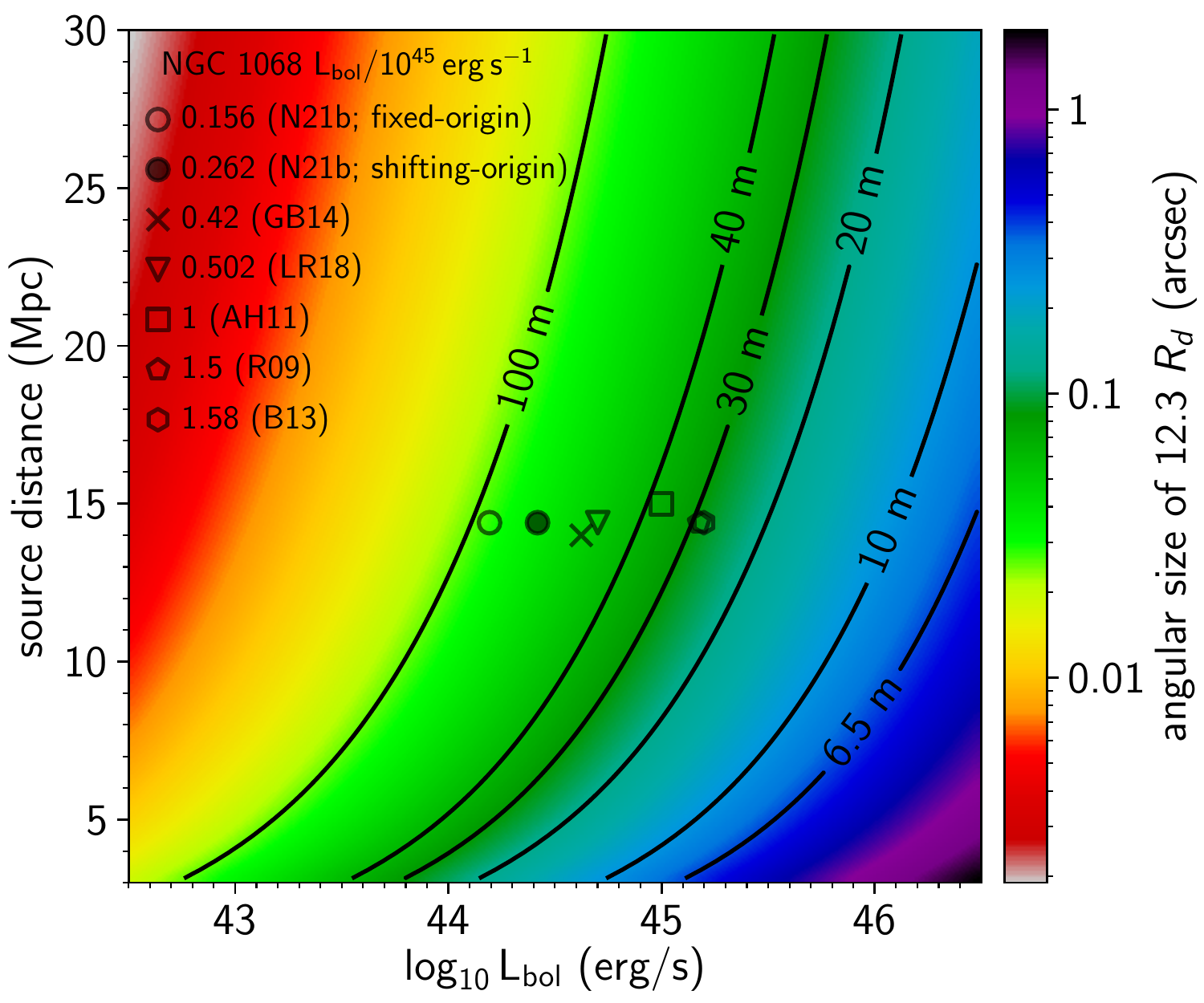}
  \caption{Angular size on the sky (in arcsec) of a structure
    12.3~\Rd\ large (corresponding to the $y$ offset between image
    origin and brightest pixel in the rightmost panel of
    Figure \ref{fig:sizes}), as a function of AGN bolometric luminosity
    and source distance, shown as a logarithmic color map. Labeled
    contour lines trace the diffraction limit at $10\!~\mic$ of
    telescopes between 6.5 and 100~m diameter. Sources with luminosity
    and distance below a certain contour produce structures that can
    be potentially resolved by the corresponding telescope. The
    symbols indicate various bolometric luminosities reported for
    NGC~1068 in the literature, which vary by a factor of several:
    this work (N21b, fitting directly the $K$-band image from
    \gravity, two models, see \papertwo),
    \citet[][GB14]{Garcia-Burillo+2014},
    \citet[][LR18]{Lopez-Rodriguez+2018},
    \citet[][AH11]{Alonso-Herrero+2011}, \citet[][R09]{Raban+2009},
    and \citet[][B13]{Burtscher+2013}. Most references adopted
    $D = 14.4$~Mpc as the source distance, except GB14 (14 Mpc)
    and AH11 (15 Mpc).}
  \label{fig:resolvingpower}
\end{figure}

This range of luminosities translates to sizes between 1.9 and 6.2~pc
for the 12.3 \Rd\ offset and, with a distance of 14.4~Mpc to NGC~1068,
between 28 and 89~mas angular extent on the sky. The larger values are
well above the diffraction limit of 30--40~m class
telescopes. Figure~\ref{fig:resolvingpower} shows the angular size of
12.3~\Rd\ on the sky, as a function of AGN bolometric luminosity
(which sets the distance of the dust sublimation radius) and of source
distance, together with the diffraction limit of telescopes with
6.5--100~m diameters $D$ ($\theta = 1.22\,\lambda / D$) at
10~\mic. The bolometric luminosities for NGC~1068 listed above are
marked in the figure. The effect described above is potentially
resolvable with 30 and 40-m class telescopes, if good astrometry can
be achieved. Smaller telescopes would struggle, unless a source were
much brighter than NGC~1068, and/or closer than it (but such sources
do not exist). The same is true if the bolometric luminosity of the
source is on the lower end of the range. Other nearby AGNs at similar
distances have larger tori, which makes the situation easier
\citep[e.g.,][]{Alonso-Herrero+2018,Alonso-Herrero+2019,Combes+2019}.

Another drawback of half-light radii is that they cannot measure
morphology sizes independently for orthogonal directions on the
sky. In the following subsection we remedy this by introducing
\emph{radii of gyration}.

\subsubsection{Radii of Gyration}
\label{sec:rg}
\noindent
A very robust way to measure image extension is via \emph{radii of
  gyration} about the $x$ and $y$-axes
\begin{equation}
  \label{eq:1}
  \Rgx = \sqrt{\mu_{20}/\mu_{00}}, \qquad \Rgy = \sqrt{\mu_{02}/\mu_{00}}.
\end{equation}
The gyration radius about an axis is the distance to a line parallel
to the axis, at which all the ``mass'' (i.e. all the pixel values)
could be concentrated without changing the second moment about that
axis. Like other quantities computed from image moments, the concept
of gyration radius is borrowed from mechanics. Figure \ref{fig:sizes}
shows the gyration radii \Rgx, \Rgy\ as the semimajor axes of a red ellipse,
plotted over the brightness distributions produced by a \C\ model. The
ellipse is centered on the image centroid $(\bar x, \bar y)$.

When estimating morphology sizes, the radii of gyration have several
benefits over other measures, such as the half-light radius. One is
that they give realistic estimates of both the $x$ and $y$ extensions
independently. This allows us then to compute the \emph{elongation} or
\emph{aspect ratio} of the brightness distribution, which we will
explore in the next subsection. Further, in all but the \hbox{\iv\ =
  90\degr} case, where the image may show two bright centers of
emission offset in the $\pm y$-direction, the ellipse made of gyration
radii and anchored on the image centroid follows the bulk of the
emission more faithfully than circular apertures pinned to the
telescope pointing (usually the suspected locus of the AGN).
\begin{figure} \center
  \includegraphics[width=\hsize]{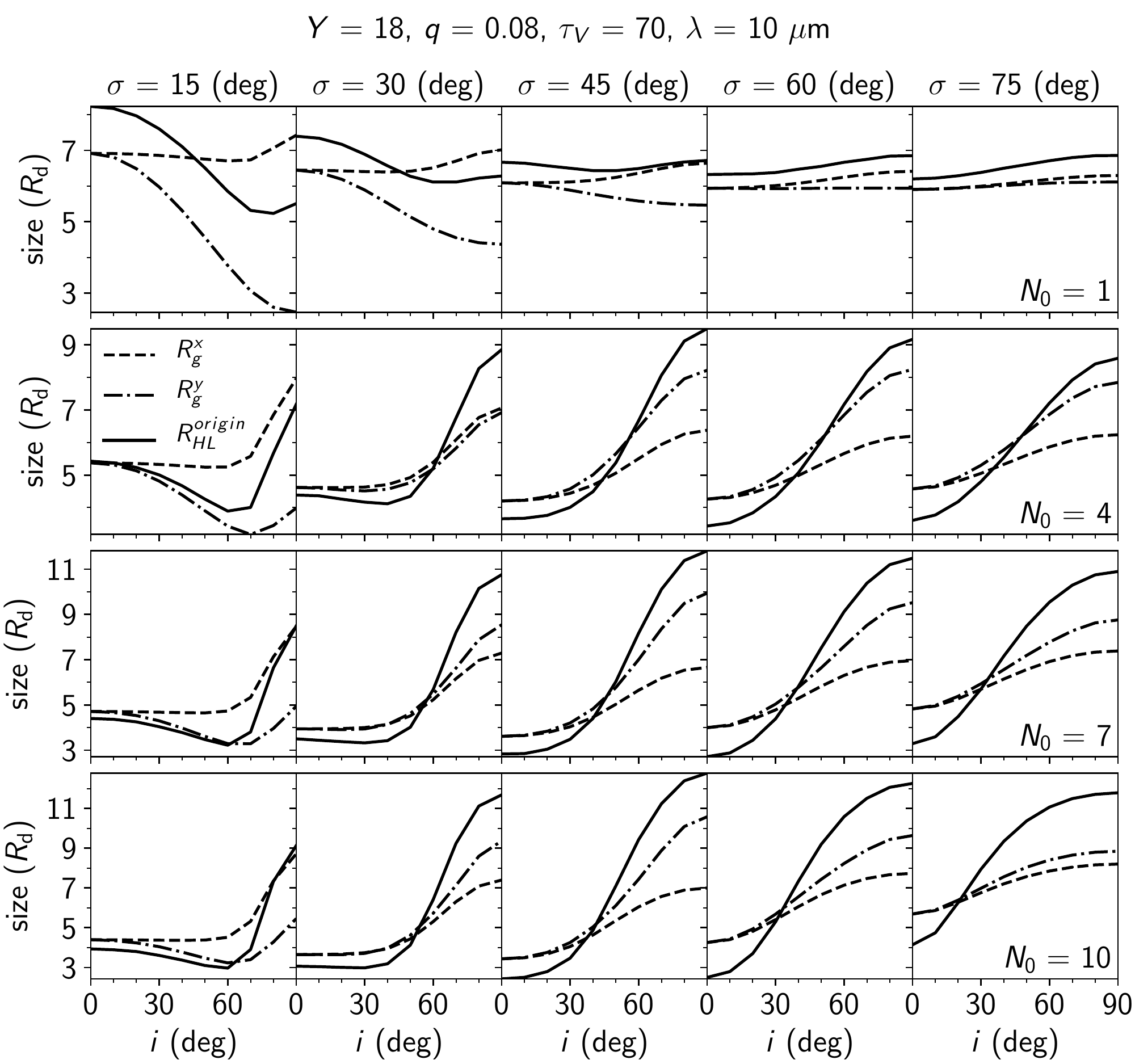}
  \caption{Morphological size measurements of brightness maps at
    10~\mic. Torus parameters $\Y = 18$, $\q = 0.08$, and $\tv = 70$
    are fixed, while \sig, \No, and \iv\ vary as indicated. Vertical
    scale is the same only per row. Solid lines show the half-light
    radius \RHorig\ of an aperture centered on the image
    origin. Dashed lines show the radius of gyration in $x$-direction,
    \Rgx, while dashed-dotted lines show \Rgy, the radius of gyration
    in $y$-direction.}
  \label{fig:rgyr_rhl}
\end{figure}

Figure~\ref{fig:rgyr_rhl} shows how the radii of gyration, \Rgx\ and
\Rgy, and the half-light radius, \RHorig, depend on (a subset of) the
torus model parameters. The analysis was
performed on the ``infinite'' resolution model images (at 10~\mic) to
illustrate direct physical effects of parameter changes on image
morphology. Plotted as a function of viewing angle, the half-light
radius varies most extremely among the three shown size
measurements. In most cases (\No\ = 1 being the exception) \RHorig\ is
the lowest at pole-on viewing angles and the largest at edge-on
orientations. That is, the half-light radius likely
underestimates/overestimates the characteristic sizes at
pole-on/edge-on viewings. The radii of gyration respond more gently to
changes in model parameters. Naturally, \Rgy\ varies more strongly than
\Rgx\ with viewing angle \iv\ and torus angular size \sig, as these two
quantities have the largest influence on the size of the image
morphology in the $y$-direction. \No\ (and \tv, which is not varied in
this figure) provides the overall magnitude of the measured size.

In the case of very high \sig\ and \No, i.e. an optically and
geometrically thick dust ``cocoon'' scenario (lower right panel in
Figure \ref{fig:rgyr_rhl}), both radii of gyration show very little
variation with viewing angle, while the half-light radius still
changes strongly between pole-on and edge-on viewing.

\subsubsection{Aspect Ratio / Elongation}
\label{sec:aspect}
\noindent
Because the radii of gyration about the $y$ and $x$-axes measure
morphological size independently in both directions, their aspect
ratio determines the elongation of the morphology
\begin{equation}
  \label{eq:elongation}
  e = \Rgy\, /\, \Rgx.
\end{equation}
In our model images the position angle is always $\pa = 0$\degr,
i.e., \Rgy\ corresponds to the polar direction and \Rgx\ to the
equatorial direction.

\begin{figure*} \center
  \includegraphics[width=0.49\hsize]{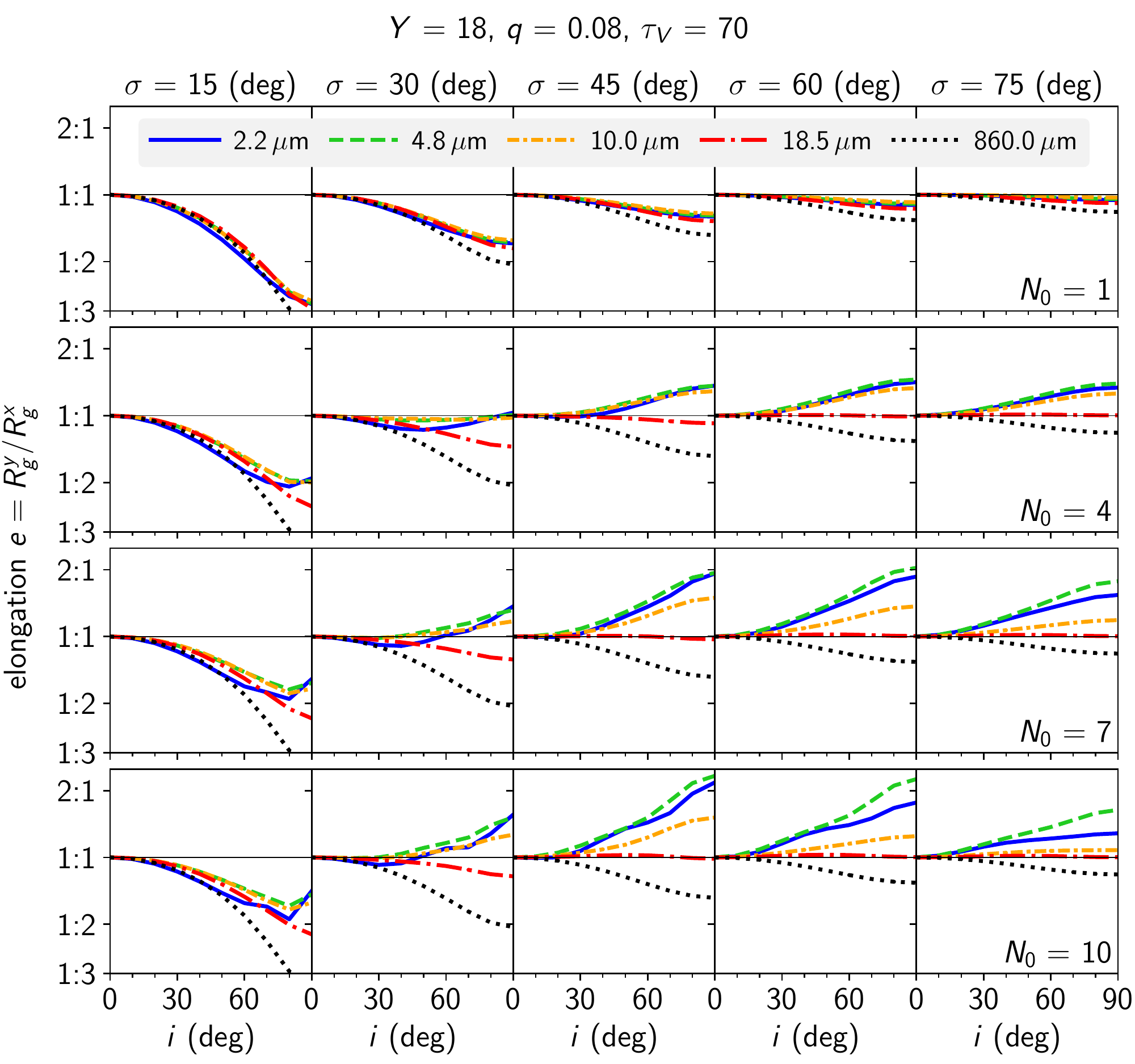}
  \includegraphics[width=0.49\hsize]{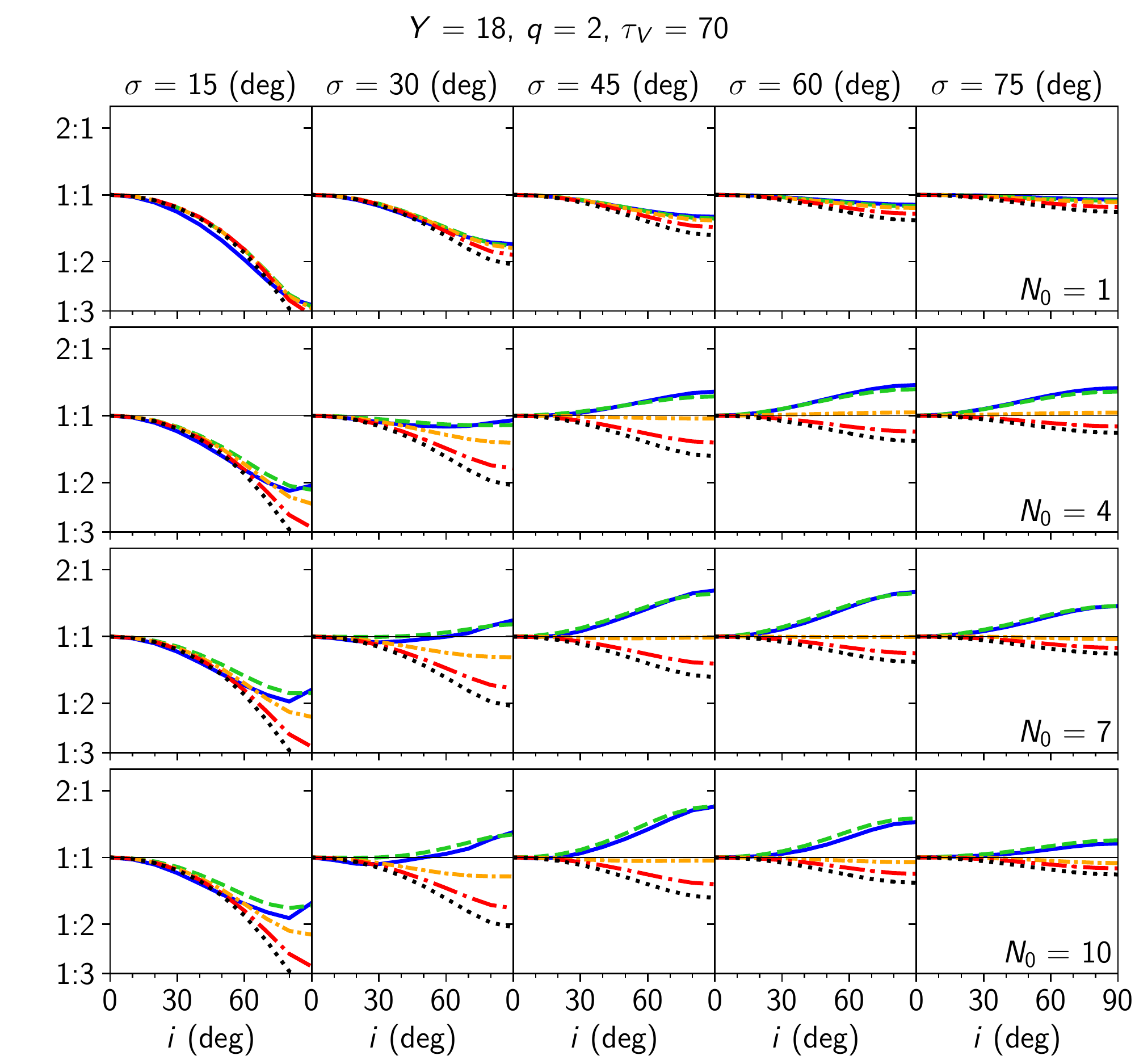}
  \caption{Elongation $e = \Rgy / \Rgx$ of image morphology. The
    vertical axis is logarithmic, which is the natural measure for
    ratios, with some integer ratios marked. A thin horizontal solid
    line in every panel marks the 1:1 aspect ratio, i.e. $e =
    1$. Values $e>1$ show elongation in the $y$-direction (polar
    extension), and $e<1$ in the $x$-direction (equatorial extension). The
    five line styles and colors encode five different wavelengths from
    NIR to submillimeter (see legend). Torus parameters $\Y = 18$ and
    $\tv = 70$ are fixed, while \sig, \No, and \iv\ vary as
    indicated. The radial steepness of the cloud distribution law is
    flat in the left panels (\q\ = 0.08) and steep in the right panels
    (\q\ = 2). For flat distributions polar elongations can be
    achieved at NIR and MIR wavelengths. At steep
    distributions no elongation in the MIR is possible.}
  \label{fig:elongation}
\end{figure*}

Using this technique, we now want to measure the elongation of torus
images produced by \C\ while varying the model
parameters. Figure~\ref{fig:elongation} shows the elongation measured
from images for a number of parameter variations, at five wavelengths
between NIR and submillimeter, and for two different radial cloud
distribution power laws (flat, with \q\ = 0.08 on the left, and steep,
with \q\ = 2 on the right). The measurements are considering all
pixels in an image unless stated otherwise.
Several characteristics are immediately apparent. All curves begin at
precisely $e = 1.0$ for \iv\ = 0\degr. This is so because for pole-on
LOSs the torus appears perfectly circular. Conversely, the largest and
smallest aspect ratios are realized at or very close to edge-on views
(\iv\ = 90\degr).

Another observation is that it is impossible to achieve any level of
polar elongation at small values of \sig\ or at the smallest values of
\No. A torus whose distribution of dust clouds is disk-like and
concentrated in the equatorial plane will always appear elongated in
the equatorial direction, i.e. perpendicular to the torus axis. At
small \q\ values no net polar elongation can be achieved at FIR and
submillimeter wavelengths with any combination of parameters. This is
because the total optical depth through the torus is small at these
wavelengths, so the emission morphology mostly follows the dust
distribution, i.e. it resembles the torus in appearance (see
Section~\ref{sec:projected_dust_maps} for more details). For steep
distributions (\q\ = 2) no polar elongation is possible even in the
$N$ band (MIR).

Significant polar elongation in emission can be easily achieved at
intermediate to large viewing angles, moderate to large \No, and
wavelengths between the $K$ and $N$ bands (for small \q), with the
highest aspect ratios observed in the $M$ band, around 4.8~\mic. The
highest elongations require \hbox{$\sig \approx 45\degr-60\degr$}, and
$\iv \approx 90$\degr.

\begin{figure}
  \center
  \includegraphics[width=\hsize]{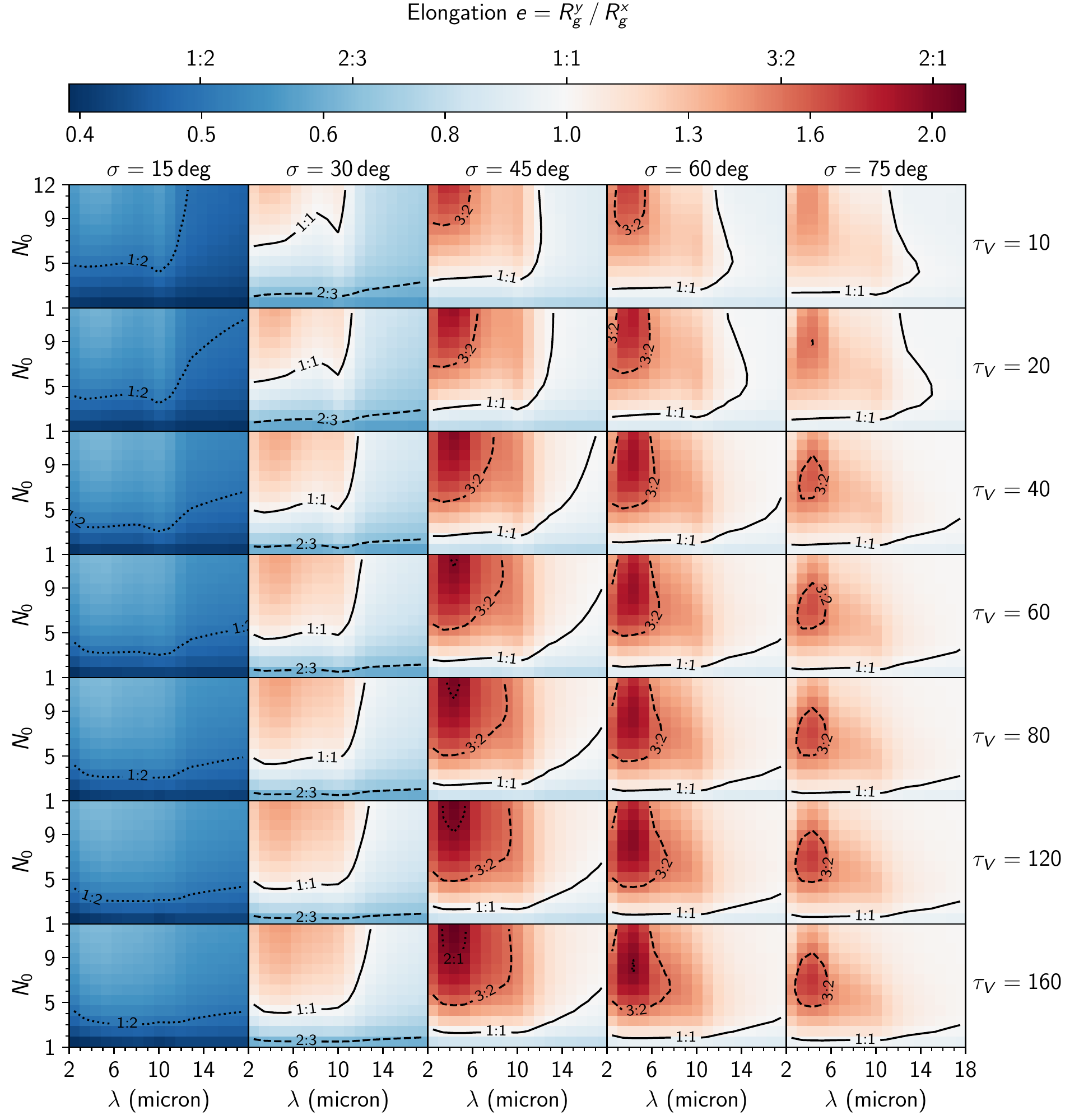}
  \caption{Image elongation $e$ as a function of model parameters, shown
    as heat maps. Red colors show elongation in the $y$-direction (polar),
    blue colors in the $x$-direction (equatorial). Some best-fit parameter
    values from SED fitting of NGC~1068 are held fixed, i.e., \iv\ =
    75\degr, \q\ = 0.08, \hbox{\Y\ = 18}, while all other parameters
    vary as indicated. All panels are normalized to the same range and
    are logarithmically stretched separately for values
    smaller/greater than 1.0 (see color bar). Contour lines show
    aspect ratios 1:1 (solid), 2:3 and 3:2 (dashed), and 2:1 and 1:2
    (dotted), also marked on the color bar (upper tickmarks).}
  \label{fig:elongation_heatmaps}
\end{figure}

Figure~\ref{fig:elongation_heatmaps} shows the aspect ratio for many
parameter variations, but as heat maps. Parameters \iv\ = 75\degr, \q\
= 0.08, and \Y\ = 18 are held fixed, and the other parameters vary as
indicated.
Regions of significant polar elongation are clearly visible as areas
of deep red color in the panels. The highest realized polar elongation
for the NGC~1068 case shown in the figure is $e = 2.13$ and is
realized at
$(\sig,\No,\tv,\lambda) = (45\degr,12,160,\about 4.0~\mic)$. The
smallest ratio in the figure, i.e., the highest in equatorial
direction, is $e = 0.39$. More extreme ratios require more edge-on
viewing angles.

MIR observations with long-baseline interferometry on VLTI
\citep[][and references therein]{Burtscher+2013} measured polar
elongations of nuclear dust emission for a number of sources: \about 2
(for Circinus), 1.3 for NGC~3783, 1.5 for NGC~424, and 1.3 for
NGC~1068, the example object of our study. These elongations are
easily achievable with \C\ torus images. For instance, with \iv\ held
at 75\degr, model image elongation ranges up to 1.5 at 10~\micron; it
can reach well above 2 at higher inclinations. Thus, \C\ torus models
naturally can reproduce the polar elongation of the spatial flux
distribution in the $N$ band up to several parsecs away from the AGN.

Interferometric observations that include shorter baselines are
sensitive to the brightness distribution on larger spatial scales,
i.e., farther away from the nucleus. \citet{Lopez-Gonzaga+2016b}
reported on such observations with VLTI, adding the AT telescopes, and
found even larger polar elongations on such scales for several
sources: 1.9 for NGC~424 (and for Circinus), 2.3 for NGC~1068, 2.5 for
NGC~5506, and 3.1 for NGC~3783. In addition, \citet{Leftley+2018} find
$e=2.9$ for ESO323-G77. Note that for Circinus and NGC~1068 these
values are lower limits, as the extended emission is over-resolved
interferometrically and continues in single-dish
observations. NGC~3783 and ESO323-G77 are type 1 AGNs, meaning that our
viewing lines to them are most likely not too highly inclined. As we
have seen in figures \ref{fig:elongation} and
\ref{fig:elongation_heatmaps}, high inclinations are necessary to
produce significant polar elongation of the MIR morphology. In these
type 1 sources a two-component model is therefore almost certainly
preferred. For the above measurements, the authors estimated polar
elongation by fitting 2D Gaussians, but the FWHM of a Gaussian is
equivalent to the radii of gyration we use here (although our
moment-based method works for any morphology).

Achieving elongations of 2.5 or even 3 with single-component models
(i.e. a torus) may be challenging, or impossible in type 1
orientations. Two-component models, such as a disk and a dusty
outflow, provide the necessary morphology by design, but at the cost
of higher model complexity. The measured elongation, however, always
depends on the pixels considered. Just as in the case of VLTI
measurements with UT baselines only, versus UT+AT baselines, the
sensitivity to larger spatial scales changes the derived
elongations. Similarly, an instrumental flux sensitivity limit will
affect which pixels will in fact be considered when estimating
morphological parameters.

\begin{figure}
  \center
  \includegraphics[width=\hsize]{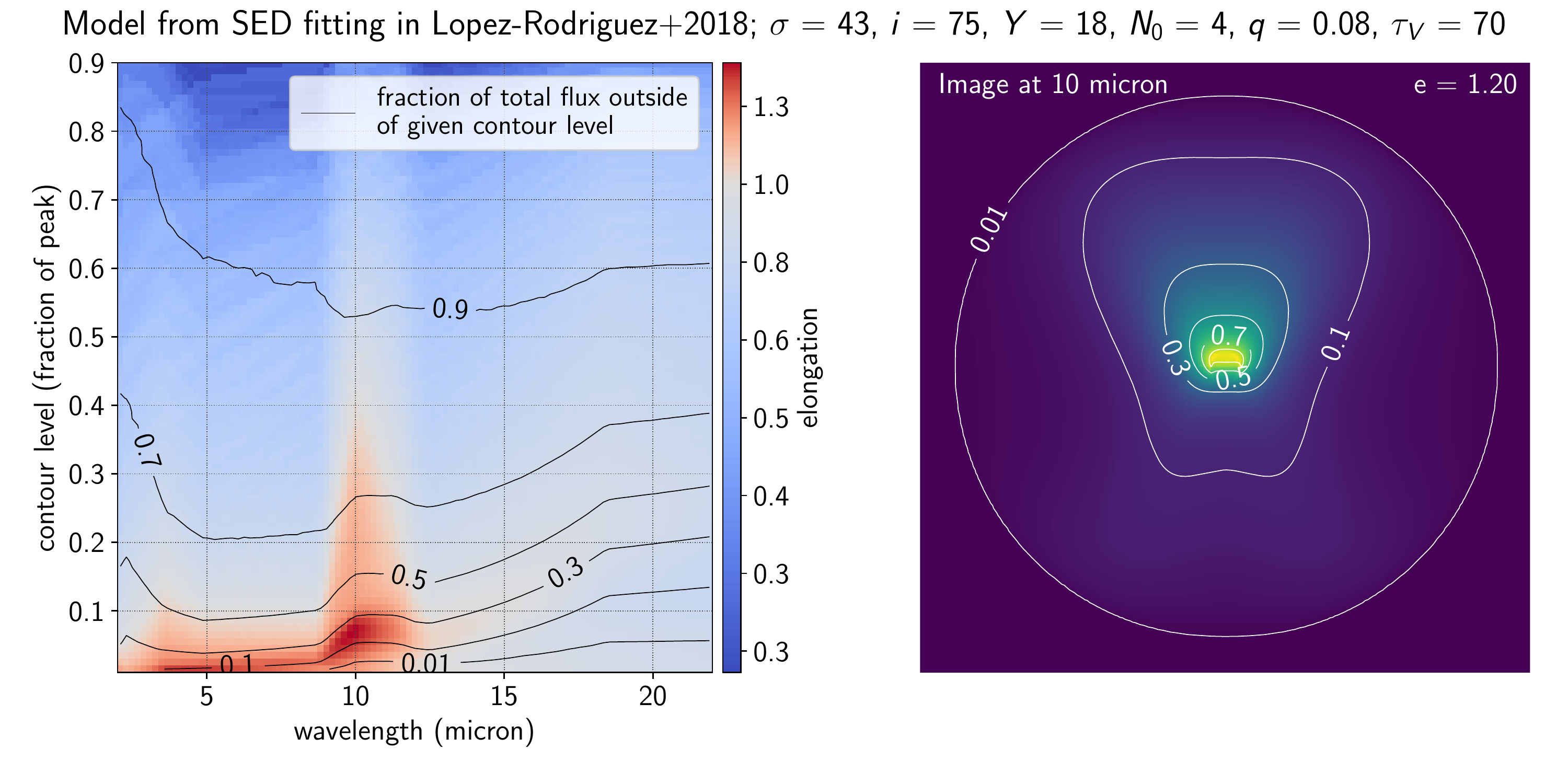}
  \includegraphics[width=\hsize]{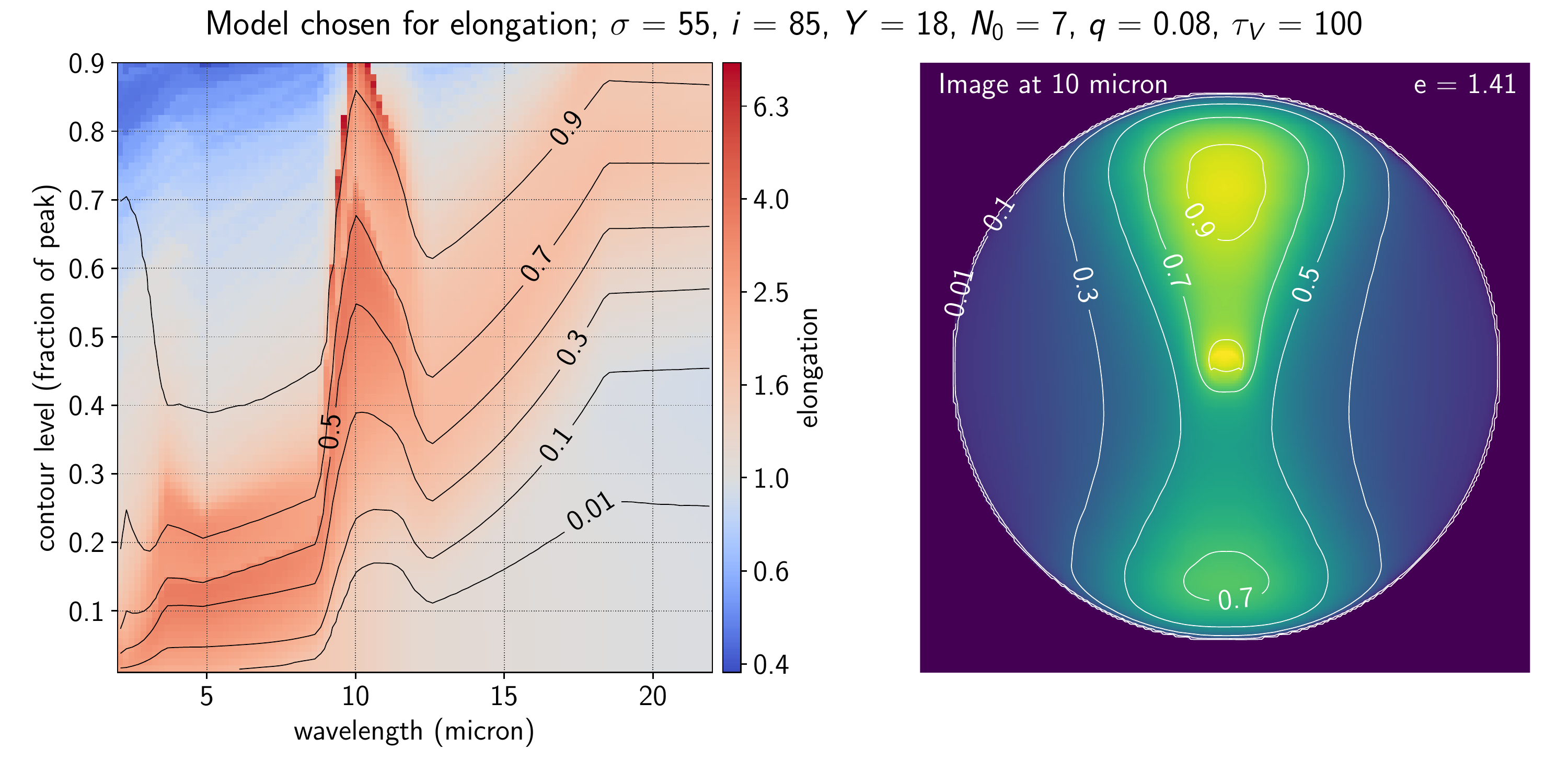}
  \caption{The effect of sensitivity on measured elongation. Two
    models are shown. Top: with parameters derived from SED
    fitting in \loro. Bottom: similar to the top model, but with
    parameters modified for elongation. Each left panel shows as a
    color map the elongation $e = \Rgy/\Rgy$ of all pixels contained
    \emph{within} a given contour level (given as a fraction of the
    peak) on the \yax\ and of wavelength on the \xax. For instance, in
    the bottom model at 10~\mic\ and 0.5 contour the elongation
    reaches 3.22, while the entire image (i.e. contour level 0) has
    only $e = 1.41$. The overplotted black lines show the fraction of
    total flux that is \emph{outside} of a given contour, i.e. is
    being missed at a given sensitivity level. For example, in the
    bottom model at 10~\mic\ and 0.5 isophote a fraction of 0.43 of
    the total flux is outside that isophote. The right panels show the
    image at 10~\mic\ from each model, with contour levels shown as
    white lines at 0.01, 0.1, 0.3, 0.5, 0.7, 0.9 of the peak. These
    contours correspond to the \yax\ in the left panel. The elongation
    $e$ printed in the upper right corners is for the \emph{entire}
    image (contour level 0).}
  \label{fig:elongation_vs_contours}
\end{figure}

Figure \ref{fig:elongation_vs_contours} illustrates this with two \C\
torus models, one with parameters from SED fitting in \loro, and a
version of it with parameters somewhat modified to increase
elongation. The elongation measured as a function of wavelength and
sensitivity level (i.e. which contour is the limiting isophote) varies
strongly. In the \loro\ model, at 10~\mic, considering the entire
image yields $e = 1.2$, but it reaches 1.43 for the 7\% contour. This
SED-derived model is elongated enough to explain the value of $e=1.3$
measured by VLTI for NGC~1068 \citep{Burtscher+2013} with long
baselines alone. It is not elongated enough to explain $e=2.3$ derived
for VLTI observations \cite{Lopez-Gonzaga+2016b} performed on larger
spatial scales.

Modifying some of the model parameters (e.g., higher \sig, \iv, \No,
\tv) can produce models with larger polar elongations. In
Figure \ref{fig:elongation_vs_contours} the bottom row shows one such
model; it produces $e=1.41$ at 10~\mic\ if the entire image is
considered. However, the same model shows much higher elongations at
other sensitivity levels, e.g., $e = 3.22$ at the 0.5 level of the
peak.

To match observations, not only the elongation but also the angular
size of the morphology must be considered. In this release of \HC\ our
sampling of the torus size (the \Y\ parameter) tops out at \Y\ =
20. With larger values and sufficiently high luminosity it is possible
that such elongations could be maintained on larger physical
scales.

\begin{figure}
  \center
  \includegraphics[width=\hsize]{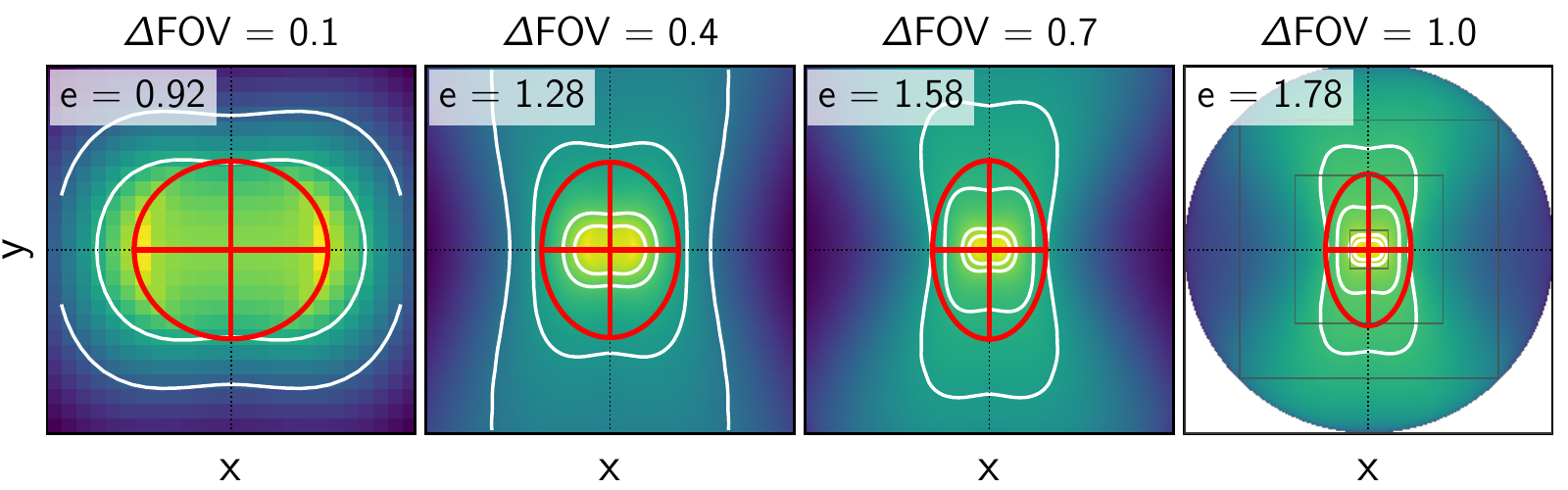}
  \caption{Image elongation as a function of FOV. The same model is
    shown in all panels, with parameters \sig\ = 45\degr, \iv\ =
    90\degr, \Y\ = 20, \No\ = 7, \q\ = 0, \tv\ = 40, $\lambda$ =
    4.8~\mic. The rightmost panel shows the full image, while the
    three left panels show fractional FOVs ($\Delta$FOV = 0.1, 0.4,
    0.7 of the full FOV). Light-gray squares in the right panel
    indicate the other FOVs. The intensity maps are log-stretched,
    while contour lines (white) are computed on the linear image at 5\%,
    10\%, 30\%, and 50\% percent of the peak pixel. Red ellipses show the
    morphological sizes measured by radii of gyration. From small to
    large FOV, the measured elongation $e = \Rgy / \Rgx$ turns from
    slightly elongated in the $x$-direction (equatorial) to very elongated
    in the $y$-direction (polar), as indicated in the panels.}
  \label{fig:elongation_vs_fov}
\end{figure}

Somewhat similar to the question of brightness level sensitivity, a
potential pitfall in resolved observations is that the measured
elongation of a brightness distribution depends on the FOV. This
effect may even invert the axis of elongation, depending on the
FOV. Figure~\ref{fig:elongation_vs_fov} demonstrates this with a
4.8~\mic\ image of a \Y\ = 20 torus model, looked at with
progressively larger FOVs. From small to large FOV, the measured
elongation $e = \Rgy / \Rgx$ turns from slightly elongated in the
equatorial direction to very elongated along the polar axis, as
indicated in the panels.

\subsubsection{Gini Coefficient as Indicator of Image Compactness}
\noindent
The Gini coefficient / index \citep{Gini1921} was originally developed
in the field of economics to characterize the relative inequality of
income distributions. It is applicable generally, though, and is a
useful measure of image compactness in our case. Applied to the
distribution of pixel values, it measures the relative deviation of
the distribution from uniformity. In the discrete case, the Gini
coefficient of an array $I$ is
\begin{equation}
  \label{eq:gini}
  G = \frac{\sum_i (2 i - n - 1)\cdot I_i}{n \sum_i I_i}
\end{equation}
where the $n$ values $I_i$ of the array are sorted in ascending order,
and $i$ are the corresponding (one-based) indices of the sorted
array. In our case of a 2D image, $I(x,y)$ can be simply flattened
prior to computing $G$. The Gini index is independent of the absolute
magnitude of the pixel values; only their relative distribution
matters. If all values are identical, $G=0$. If the distribution is
randomly uniform, $G=1/3$. In the most unequal distribution, a delta
function (e.g., when exactly 1 pixel has nonzero value), $G = 1$.

\begin{figure}
  \center
  \includegraphics[width=\hsize]{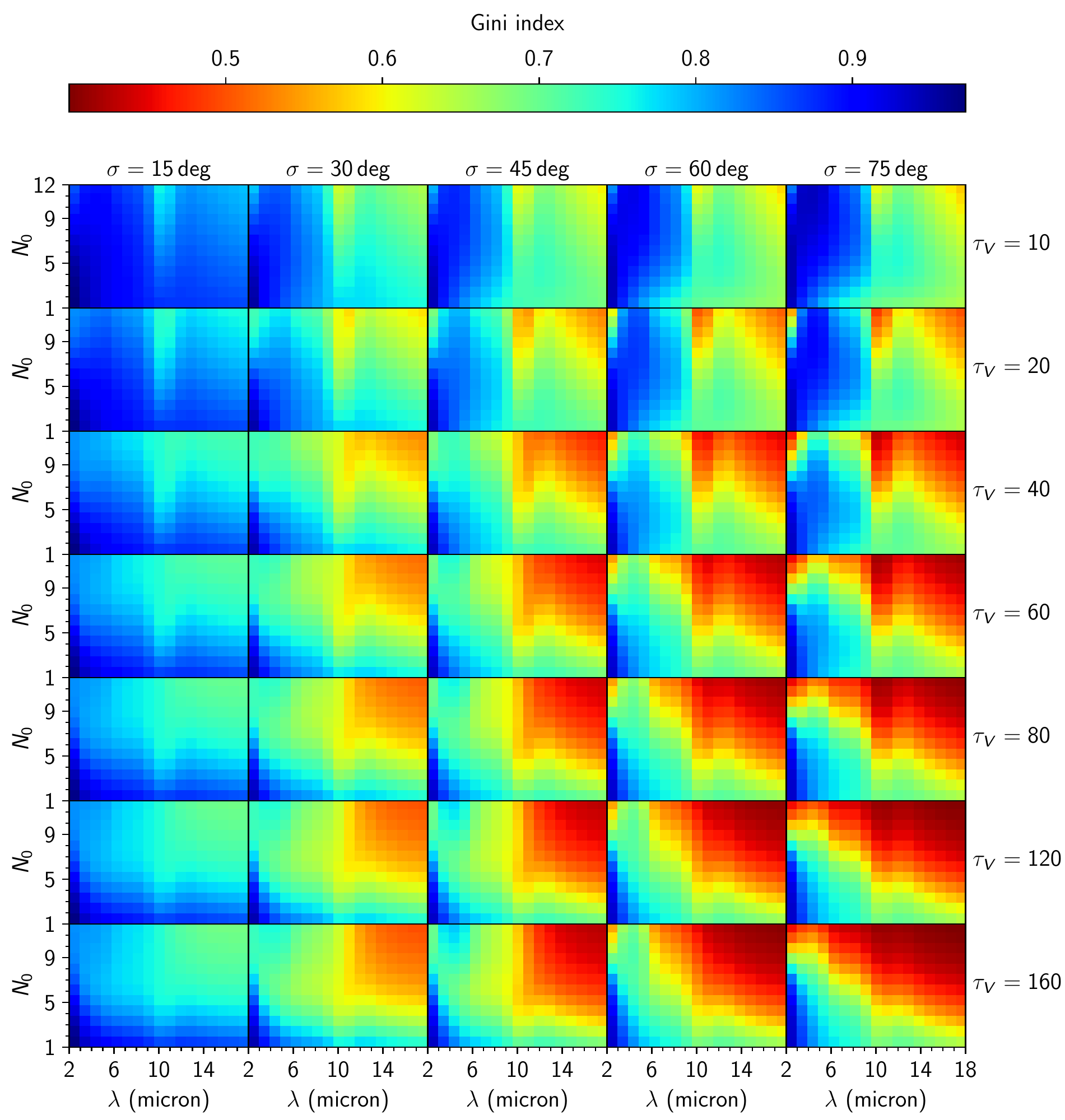}
  \caption{Image compactness, measured as the Gini index, for a number
    of model parameters that vary as noted. $Y=18$, $q=0$ and
    $\iv = 90$\degr\ are fixed. Blue colors are compact morphologies,
    red colors are extended images. The maximal range of the Gini
    index is from 0 (all pixels equal) to 1 (a single pixel is
    nonzero). The size enhancement around 10.0~\mic\ due to the
    increased dust absorption coefficient is clearly visible.}
  \label{fig:gini}
\end{figure}

Figure \ref{fig:gini} shows $G$ as a function of several parameters that
vary as indicated. $Y=18$, $q=0$ and $\iv = 90$~\deg\ are fixed.
Values close to unity (dark blue colors) are the most compact
morphologies, where all flux is concentrated within a small fraction
of all pixels. Lower values (red colors) represent broader
distributions. For the $Y=18$ torus embedded in a
\hbox{$(2\,Y_{\rm max})^2 = 40 \times 40$}~FOV, the smallest
achievable $G$ value is \about 0.36. This would be the case if all the
pixels within a $Y = 18$ circle had the same value while pixels
outside that circle were zero (we are ignoring the central dust-free
cavity for the sake of simplicity). The smallest $G$ in
figure~\ref{fig:gini} is 0.40, i.e. there are models whose appearance
approaches that of a uniformly bright disk. Not surprisingly, the
corresponding model parameters represent the thickest and most dense
torus of all shown in the figure, namely
$(\sig,\No,\tv,\lambda) = (75~\deg, 12, 160, 18~\mic)$. The elongation
of this morphology is $e = \Rgy / \Rgx = 1$. On the other hand, the
most compact emission morphology is realized with the most shallow
torus, smallest number of clouds, lowest optical depth, and shortest
wavelength $(\sig, \No, \tv, \lambda) = (15\,\deg, 1, 10,
2\,\mic)$. This smallest model has $G = 0.97$ and elongation
$e = 0.31$. Figure~\ref{fig:minmaxtorus} in
Appendix~\ref{sec:appendix-gini} shows both extremes.

In general, compact images are achieved at small \No, small $\lambda$,
and preferentially at small optical depths. For large optical depths,
only the very smallest \No\ and shortest wavelengths can produce
compact emission sizes, i.e. when only the hottest dust close to the
center is visible. In Figure \ref{fig:gini}, at small to moderate \tv\
and moderate to high \No\ a size enhancement is clearly visible around
10~\mic. This is because the dust cross section peaks locally around
10~\mic, increasing the optical depth, thus intercepting more photons,
also at greater distances, before they can escape the torus. At
wavelengths slightly away from the peak of the dust cross-section
curve the photon escape probability is significantly higher \citep[see
Figure 2 in][]{Nenkova+2008b}.

\subsection{Image Asymmetry}
\label{sec:asymmetry}
\noindent
The interplay between torus angular width \sig, viewing angle \iv, and
total optical depth along an LOS can shift the region that emits light
(as viewed by an observer) up or down from the image center. This
asymmetry in the distribution of emission can be quantified in several
ways.

\subsubsection{Thermal Emission Distribution}
\begin{figure} \center
  \includegraphics[width=\hsize]{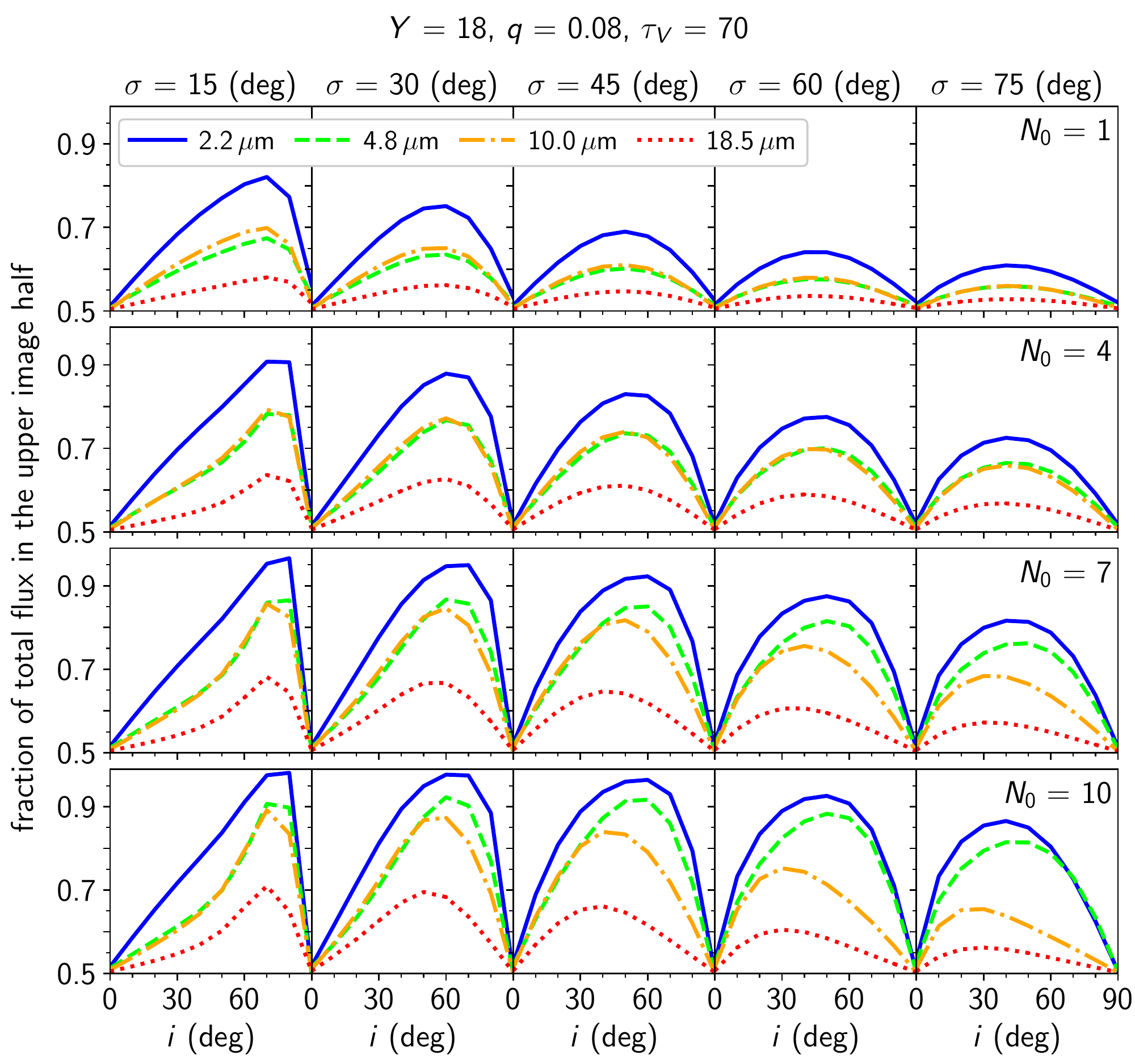}
  \caption{Fraction of total light contained in the upper image half,
    as a function of model parameters. Torus parameters $\Y = 18$,
    $\q = 0.08$, and $\tv = 70$ are fixed, while \sig, \No, and \iv\
    vary as indicated. The four line styles and colors show four
    different wavelengths (see legend).}
  \label{fig:fluxdistribution}
\end{figure}

\noindent
Probably the simplest measure of asymmetry with respect to the \xax\
is the fraction of total light contained in the upper image half. This
fraction is by design 0.5 in pole-on and edge-in views, but for other
inclination angles will depend on some of the model
parameters. Figure~\ref{fig:fluxdistribution} shows this for a number
of typical model parameter combinations and four wavelengths between
2.2 and 18.5~\mic.

The asymmetry is largest at short wavelengths ($K$ band), as this is
mostly scattered light from the far inner side of the torus. Longer
wavelengths can penetrate deeper into the torus in all directions,
diminishing the asymmetry in light distribution among the upper and
lower image halves. The fraction as a function of viewing angle always
has a clear peak, and for all $\sig>45$\degr\ and short wavelengths it
is approximately symmetric around $\iv \approx 45$\degr. At smaller
\sig\ or longer wavelengths the peak occurs roughly at $90\degr-\sig$,
i.e. at the torus half-opening angle. Increasing optical depth via
higher \No\ increases the overall scale of the emission asymmetry.

The fraction of emission from regions above the nucleus can easily
reach 90\% or more at 2.2 and 4.8~\mic, and over 80\% at 10~\mic. At
18.5~\mic\ the asymmetry tops out at \about 70\%, but it is
significantly less in most cases.

In Section \ref{sec:projected_dust_maps} we will demonstrate that
although the distribution of the thermal emission can be highly
asymmetric in many configurations, it does not necessarily correlate
with the physical distribution of the dust in the torus, whose bulk is
still in the equatorial plane.

\subsubsection{Centroid Offset}
\begin{figure} \center
  \includegraphics[width=\hsize]{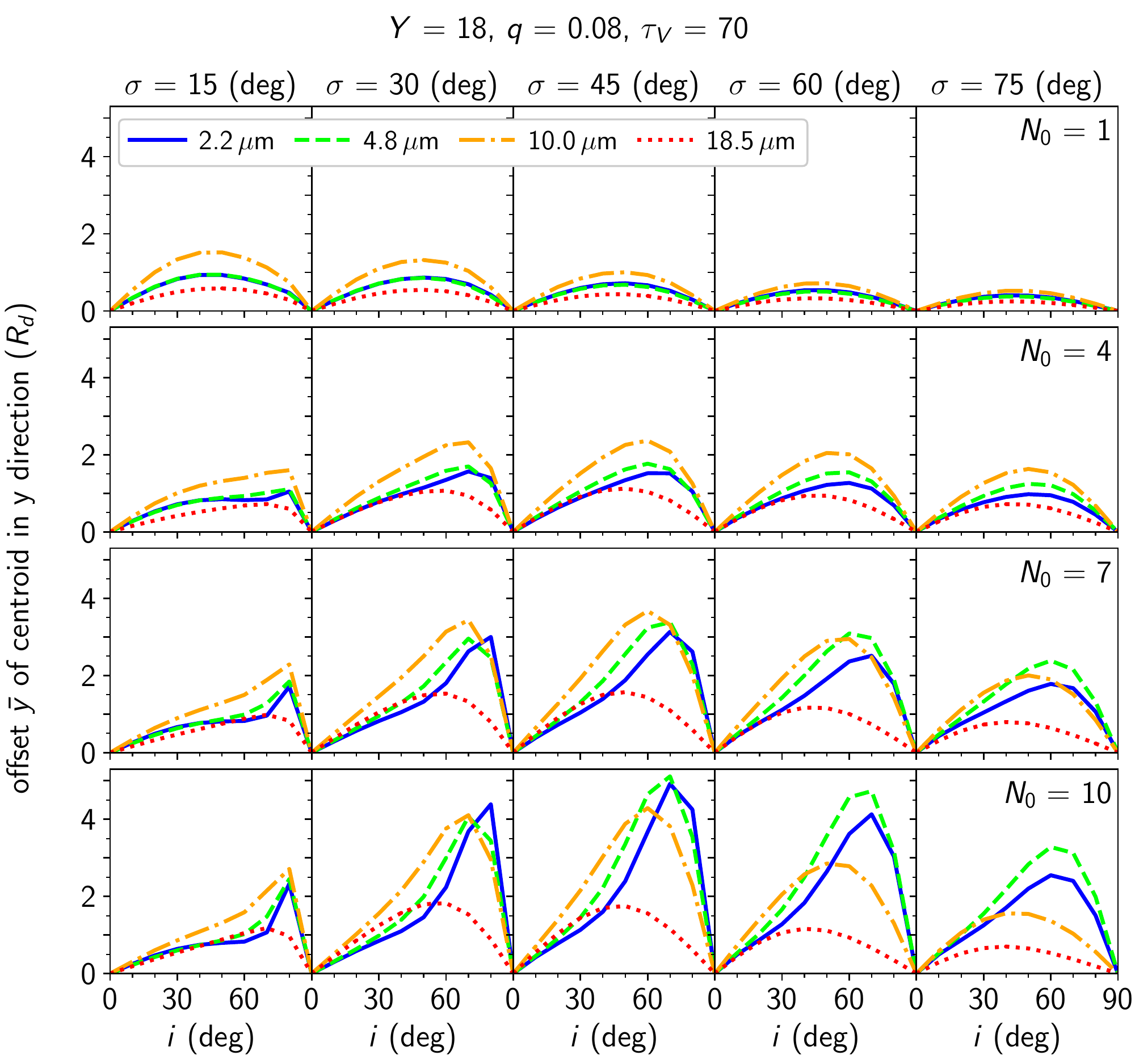}
  \caption{Vertical offset of the image centroid, $\bar y$, in units
    of dust sublimation radius \Rd. Torus parameters $\Y = 18$,
    $\q = 0.08$, and $\tv = 70$ are fixed, while \sig, \No, and \iv\
    vary as indicated. The four line styles and colors show four
    different wavelengths (see legend).}
  \label{fig:ybar}
\end{figure}

\noindent
A different method to quantify the asymmetry in emission distribution
is by measuring the $y$ position of the brightness centroid, $\bar
y$. Figure~\ref{fig:ybar} shows this in a similar fashion to
Figure~\ref{fig:fluxdistribution}, but the \yax\ in each panel is the
vertical offset of the light centroid from $y$=0 in units of dust
sublimation radius \Rd.
In $K$ through $M$ bands these curves always peak close to edge-on
viewings, but not quite at $\iv = 90$\degr\ of course, as this is a
symmetric morphology. The N band behaves similarly, albeit
transitioning to the longer-wavelength case, where the $Q$ band curve
shows a much more symmetric shape around $\iv \approx 45$\degr. The
2.2 and 4.8~\mic\ images show the largest centroid shifts (of very
similar magnitude) and can be quite substantial, up to 5~\Rd, i.e.,
almost 30\% of the torus overall radius in this $\Y=18~\Rd$ case. The
centroid offsets at 10.0~\mic\ are a bit smaller, and at 18.5~\mic\
they are the smallest of all. Again, this is due to long-wavelength
photons traveling more freely in all directions and into the torus,
thus reducing the asymmetry. The dependency of $\bar y$ on \No\ is
stronger than that of the emission fraction in the upper image half
discussed in the previous subsection.

\subsubsection{Skewness}
\noindent
Using the image moments developed earlier, the asymmetry of emission
distribution can be quantified by measuring skewness. The skewness of
a distribution of pixel values, projected onto the axes, is a
third-order moments function:
\begin{equation}
  S_x = \mu_{03} / \sqrt{\mu_{02}^3}, \qquad S_y = \mu_{30} / \sqrt{\mu_{20}^3}.
  \label{eq:skew}
\end{equation}
It measures the deviation from a symmetric distribution in the given
direction. Since \C\ images are always symmetric about the \yax, the
skewness $S_y$ is always zero. We thus only consider $S_x$,
the image skewness about the \xax\ (i.e. in the
$y$-direction). \emph{Negative} values of skewness indicate asymmetry
toward the \emph{positive} direction of an axis. Note that a
zero-valued skewness does not guarantee symmetry, but a symmetric
distribution will always have skewness zero.

\begin{figure} \center
  \includegraphics[width=\hsize]{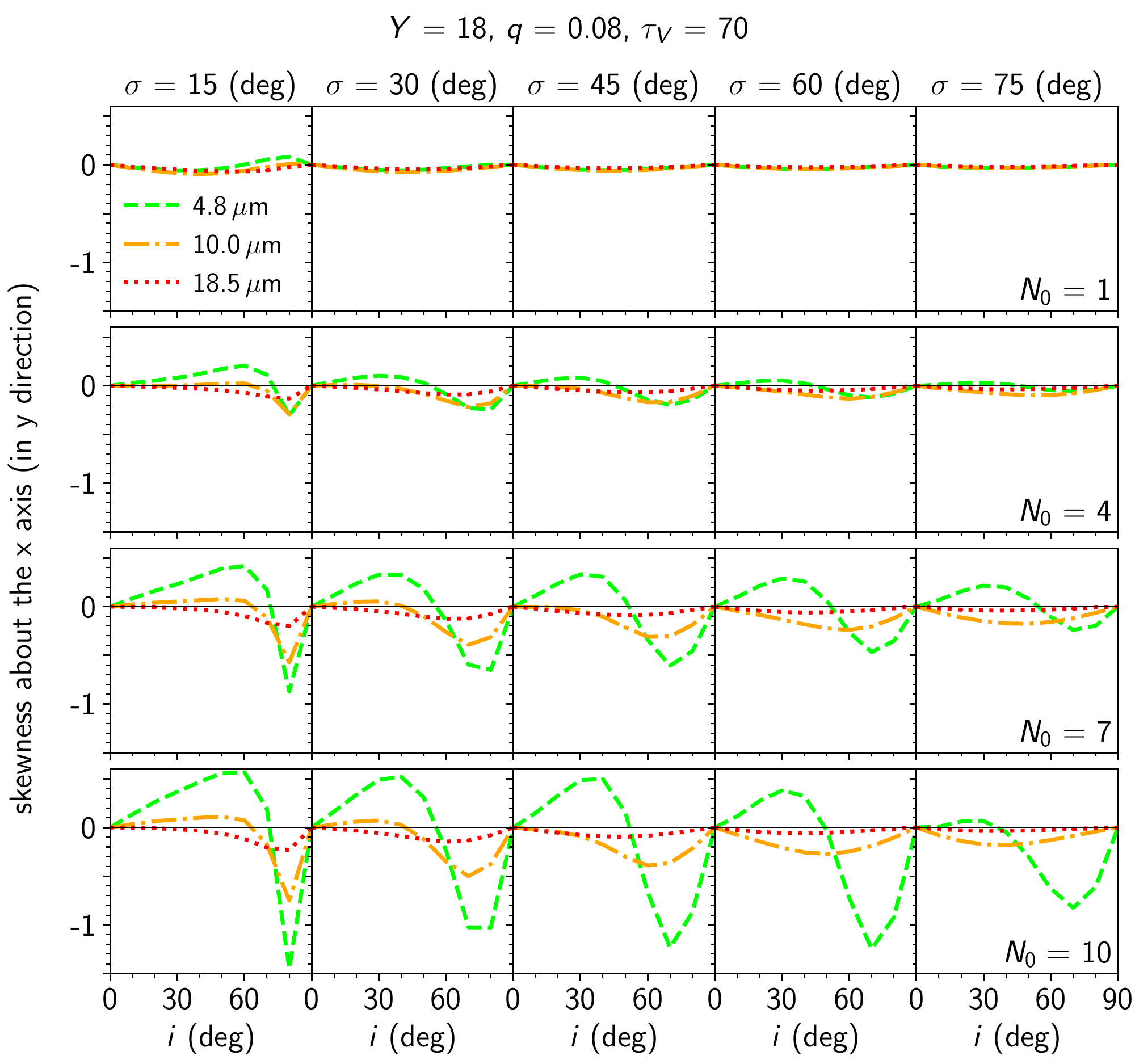}
  \caption{Image skewness about the \xax, i.e. in the
    $y$-direction. Values $>0$ indicate skewness toward the negative
    \yax, and values $<0$ indicate skewness toward the positive
    \yax. Torus parameters $\Y = 18$, $\q = 0.08$, and $\tv = 70$ are
    fixed, while \sig, \No, and \iv\ vary as indicated. The three line
    styles and colors encode three different wavelengths (see
    legend).}
  \label{fig:skewness}
\end{figure}

Figure \ref{fig:skewness} plots $S_x$ for \C\ model images at
three wavelengths and as a function of several torus model parameters
that vary as indicated. For all but the very smallest value of \No,
the skewness curves appear very similar in shape for a given \sig\ and
wavelength. The overall amplitude of the curve depends very strongly
on \No. The wavelength determines the degree of oscillations in the
curves, with the shortest wavelength showing the most variation with
viewing angle (we did not plot the curve for the $K$ band since its
range is much larger than that of longer wavelengths). At short
wavelengths (4.8~\mic) the skewness curve increases first with growing
viewing angle \iv\ (i.e. the skewness is toward the \emph{negative}
\yax). This is due to the lateral ``horns'' of the dust-free cavity in
the brightness map protruding below the image \xax. As the viewing
angle continues to increase, the curve then quickly plunges to
negative values, indicating skewness of the image toward the
\emph{positive} \yax. The viewing angle \iv\ at which this turnover
occurs depends on the torus angular width \sig. Around 10~\mic\ very
little evidence remains of positive skewness values, and the negative
excursions are of smaller amplitude than at 4.8~\mic. At long
wavelengths (18.5~\mic\ and above) almost no skewness is perceivable,
except at the smallest \sig; the images have almost no skew,
regardless of viewing angle.

\subsection{Morphologies of Projected Dust Maps}
\label{sec:projected_dust_maps}
\noindent
\begin{figure*} \center
  \includegraphics[width=\hsize]{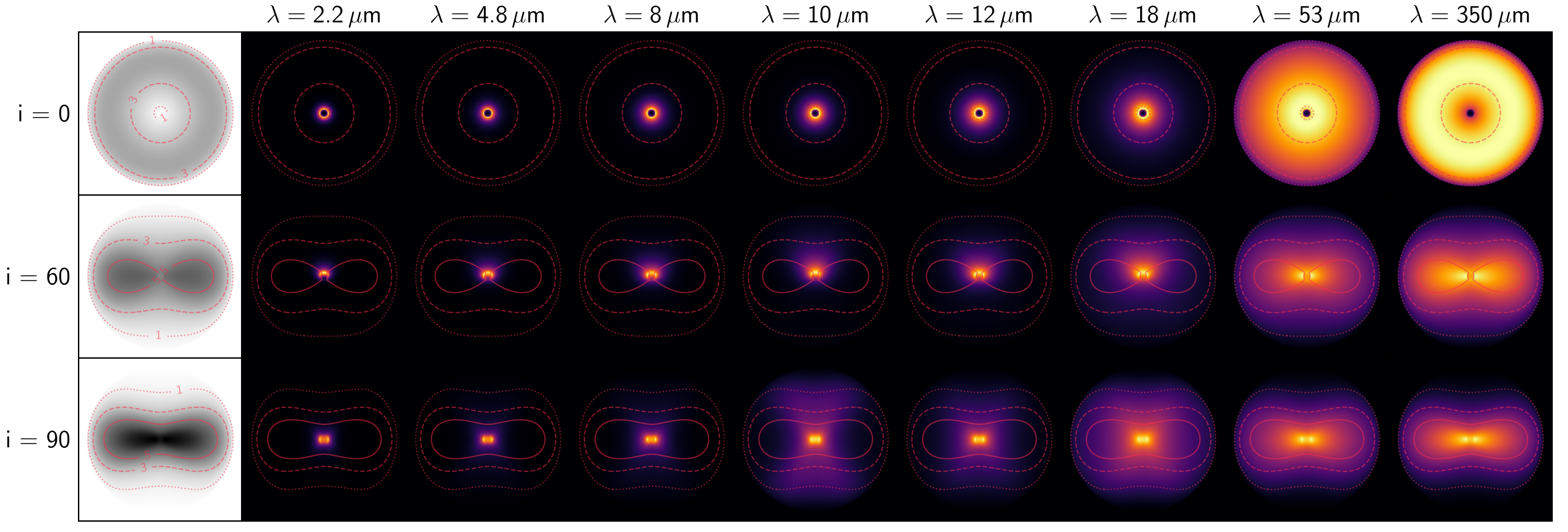}
  \caption{Torus emission maps for a model with parameters \sig\ =
    43\degr, \Y\ = 18, \q\ = 0.08, \tv\ = 70, and \No\ = 4, shown for
    three viewing angles \iv\ = 0\degr, 60\degr, and 90\degr\ (as
    rows) and for eight wavelengths between 2.2 and 350~\mic\ (as
    columns). All maps are stretched individually and linearly. The
    leftmost column shows \N, the number of clouds along an LOS,
    normalized to the global maximum $\N_{\rm max}=8.54$ across all
    viewing angles (which occurs in edge-on view). Red dotted, dashed,
    and solid contours outline one, three, and five clouds along an
    LOS, respectively. The same contours are plotted over the emission
    maps in a given row.}
  \label{fig:lightdustmaps}
\end{figure*}
The \HC\ model cubes also contain 2D maps of the projected dust cloud
distribution in the plane of the sky, $C_{d}(x,y)$. These are maps of
the number of clouds along each LOS (pixel). Whenever the LOSs through
the torus are optically thick, the morphology of the torus emission
cannot be expected to follow the dust cloud
distribution. Figure~\ref{fig:lightdustmaps} shows this clearly for
one model, at three viewing angles and a number of wavelengths. The
leftmost column shows the underlying dust cloud distribution, with
labeled contours of constant number of clouds along an LOS
overplotted. Only at FIR wavelengths above $\approx\!18~\mic$ does the
torus become globally optically thin, and the dust emission
morphologies densely fill out the cloud number contours.

The methods developed earlier for brightness maps can be used to
quantify the dust morphologies. We remind the reader that (i) the dust
maps do not depend on \tv; (ii) they depend on \No\ only as a
multiplicative factor, i.e.,
$C_{d}(\No) = \No\! \times C_{d}(\No\!=\!1)$; and (iii) moment-based
morphological measurements on these dust maps, such as radii of
gyration and elongations, are independent of \No\ as well. They only
depend on \sig, \iv, \Y, and \q, i.e. on the truly \emph{geometrical}
parameters. We can thus compare the extension of the thermal emission
maps with the extension of the dust distribution map.
\begin{figure} \center
  \includegraphics[width=\hsize]{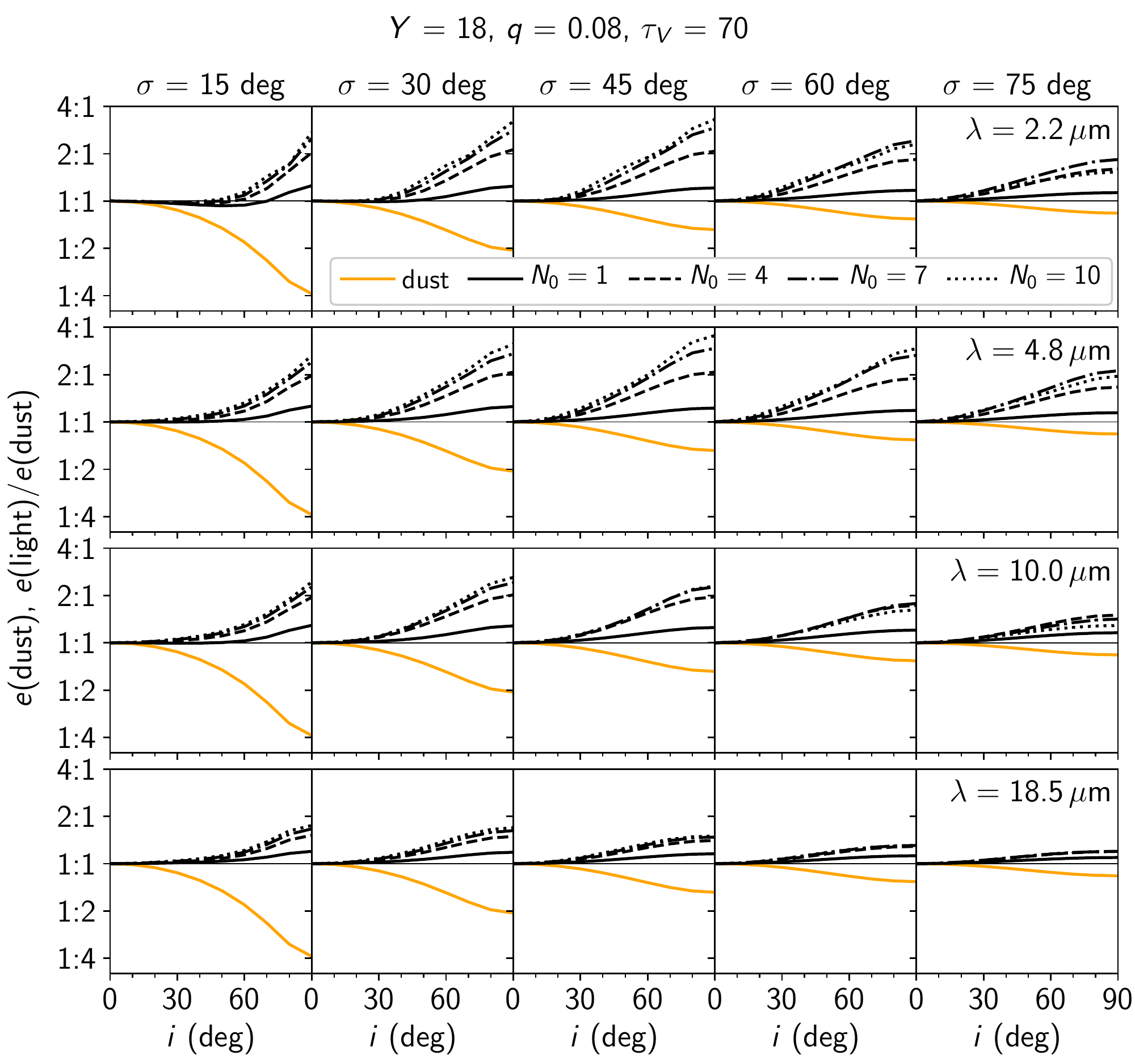}
  \caption{Elongation of the dust distribution, $e\rm(dust)$ (orange
    solid line below the 1:1 ratio), and the ratio of emission
    elongation to this dust elongation for a number of \No\ values
    (black lines; see legend). Positive values indicate that the
    elongation of the emission is more polar than the elongation of
    the dust distribution. Torus parameters $\Y = 18$, $\q = 0.08$,
    and $\tv = 70$ are fixed, while \sig, \iv\ and $\lambda$ vary as
    indicated. The vertical axis is logarithmic. Values $>1$ show
    elongation in the $y$-direction (polar extension), and $<1$ show
    elongation in the $x$-direction (equatorial extension).}
  \label{fig:light2dust_lines}
\end{figure}
Figure~\ref{fig:light2dust_lines} shows as orange lines the
morphological elongation of the 2D dust maps for a number of model
parameter combinations. \sig\ (columns) and $\lambda$ (rows) vary from
panel to panel. In every column the orange curve is the same, since
the dust distribution does not depend on wavelength. In all cases the
elongation of the dust distribution is in the equatorial direction
(i.e. ratios smaller than 1:1), which is expected given the range of
\sig\ values sampled.

The other lines show the ratio of this elongation of the emission to
the elongation of the underlying dust morphology, at four different
values of \No. In almost all cases the elongation of the emission
distribution is more polar than the elongation of the associated dust
distribution. The only marginal exception is at the smallest \sig\ and
smallest \No. The emission-to-dust elongation ratios are higher at
shorter wavelengths. At large \sig\ and wavelengths longer than \about
12~\mic\ the elongations of the emission and dust morphologies are
quite close to each other.

\begin{figure*} \center
  \includegraphics[width=\hsize]{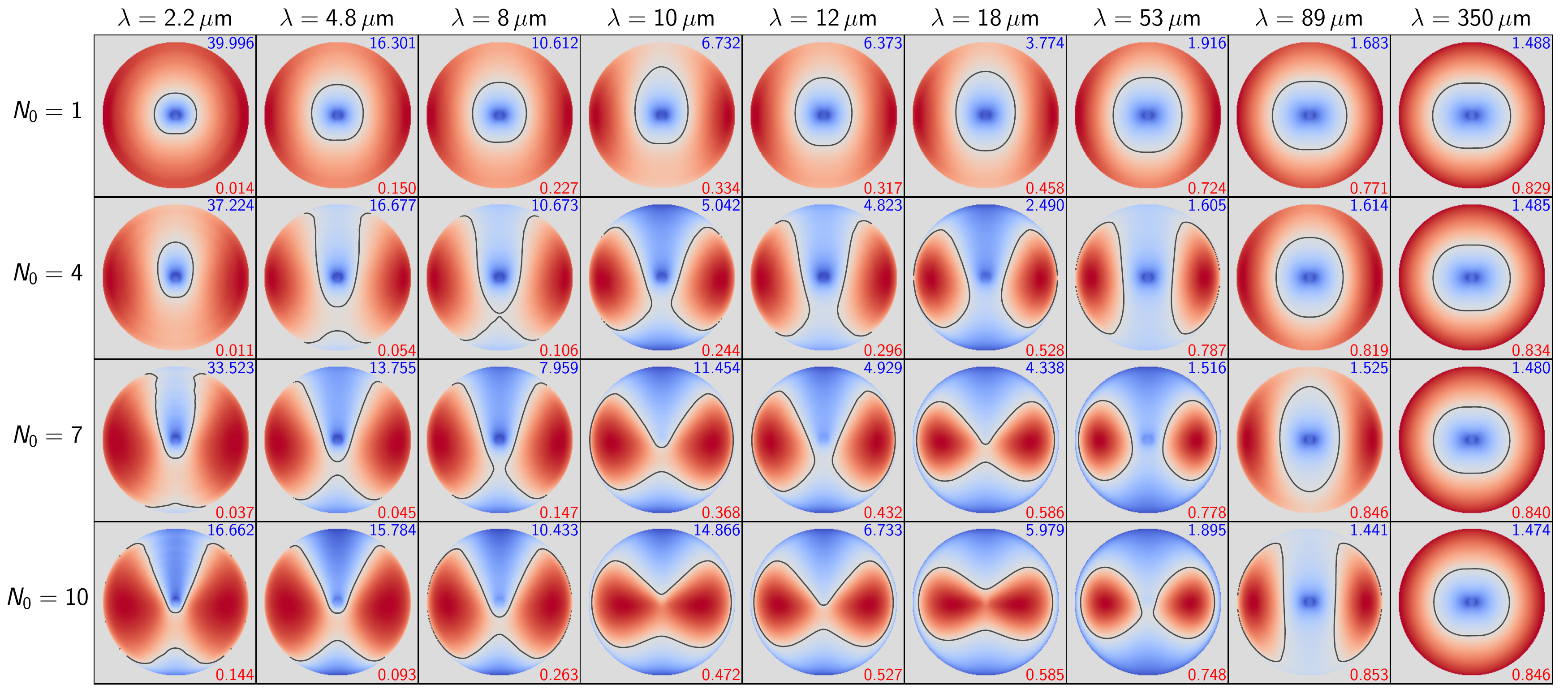}
  \caption{Ratio of emission map to dust cloud map. Regions where the
    fractional light emission dominates over the fractional dust
    content are blue, and regions where there is fractionally more
    dust present than emission produced are red. A single black
    contour line divides them. Both the input emission and dust maps
    were first normalized to unit sum, i.e. they are fractional maps
    of the total emission and total dust cloud number. Torus
    parameters $\sig = 43$\degr, $\iv = 75$\degr, $\Y = 18$,
    $\q = 0.08$, and $\tv = 70$ are fixed, while \No\ and $\lambda$
    vary as indicated. The dust cloud map is the same for every panel
    since its morphology does not depend on \No\ or $\lambda$. The
    resulting light-to-dust ratio maps are scaled logarithmically, and
    the ranges above/below zero are stretched independently
    (symlog). The dynamic range is different in each panel; the
    min/max (linear) ratios are printed in the right lower/upper
    corners in red/blue.}
  \label{fig:light2dust_maps}
\end{figure*}

Figure~\ref{fig:light2dust_maps} demonstrates spatial 2D maps of the
ratio of local emission to dust content.
The dust map is the same for all panels and corresponds to the
best-fit NGC~1068 model as before, with $\sig = 43$\degr,
$\iv = 75$\degr, $\Y = 18$, $\q = 0.08$, and $\tv = 70$. \No\ and
$\lambda$ vary between panels as indicated. Each panel shows the ratio
(computed per pixel) of the light emission map and the generating dust
map, which both were first normalized to unit sum. That is, both are
fractional maps of the total emission and total dust cloud number. The
scaling of the ratio maps is symmetrically logarithmic. A single black
contour line divides regions where the fractional emission dominates
over the fractional dust content (blue colors) from those regions
where there is fractionally more dust present than light produced (red
colors).

These maps give quick insight of the interplay between the global
morphology of photons generated by the distribution of dust. For our
model with viewing angle \iv\ = 75\degr, i.e., tilted by 15\degr\
toward the observer, it is immediately clear that at MIR wavelength
the emission has the upper hand over dust contribution in regions high
above the equatorial plane of the torus, along the system axis. When
\No\ is very small, at nearly all wavelengths the emission is
concentrated to a small circular region around the image center. The
same is true at any value of \No\ when the wavelength is very long
(submillimeter).


\section{Summary}
\label{sec:summary}
\noindent
To date only VLTI and ALMA, both interferometers, have been able to
spatially resolve the IR emission from nuclear dusty environments in
some nearby AGNs. In the future a number of extremely large single-dish
telescopes will be pointing their 24--39~m diameter (\GMT, \TMT, \ELT)
primary mirrors at AGNs as well. They will resolve the IR emission in
these sources, providing answers about the morphologies of light
emission and dust distribution that are free from model
assumptions. Resolved and model-free imagery is the only technique
capable of breaking degeneracies intrinsic to IR radiative transfer
that can produce similar SEDs from different dust geometries.

A major observational finding in recent years was that in several
nearby AGNs the MIR emission resolved with VLTI appears elongated along
the polar direction of the system. That is, the MIR light emanates
from regions above the equatorial plane of the AGN torus. Several
attempts at explaining these observations were proposed, some of which
significantly deviate from simple AGN unification models.

We developed techniques based on image moments to quantify image
morphology, and we show that thermal emission maps produced by simple
``classical'' toroidal models such as \C\ can generate significant
polar elongation in the MIR. Typical polar-to-equatorial size ratios
of up to 2:1 found by interferometric observations on physical scales
of several parsec are easily achievable by \C\ torus models and can be
even larger depending on sensitivity levels. We constrain the ranges
of parameters that produce such elongations and find that they are
rather standard. A slight tilt of the torus system axis toward the
observer is enough to produce elongations compatible with many
observations. We further find that it is impossible to obtain any
polar elongation at small values of \sig\ and \iv, at smallest values of
\No, or at wavelengths longer than \about 14--18~\mic. If the radial
cloud distribution is compact (higher \q\ values), polar elongations
are only achievable up to \about 5~\mic.

While the clumpy distributions that we model here fundamentally
provide the optically and geometrically thick material that AGN
unification calls for, and alone can account for much of the observed
polar elongation, additional components may still be present,
especially on larger physical scales (tens of pc). For example,
extended winds \citep{Hoenig_Kishimoto_2017, Izumi+2018, Asmus_2019}
and photoionization of existing material may also contribute,
including at MIR wavelengths. In some sources, such as NGC~1068, MIR
interferometric data suggest that there are additional components,
e.g., from a putative dusty outflow, which are extended on larger
scales, or even disassociated \citep{Lopez-Gonzaga+2014}; outflows on
even larger scales, and crossing to the galactic domain, are seen in
submillimeter, molecular gas observations \citep[e.g.,][and references
therein]{Aalto+2020}. These very extended components may not be
explainable with a torus model alone, and their physical connection
with the torus, or lack thereof, needs to be studied.

Evidence from radiation-hydrodynamical simulations also strongly
suggests that two-component models may be preferable, especially on
larger scales, where outflowing winds and ``failed fountains'' may
provide a good fraction of the observed MIR emission on scales of tens
of parsecs and be elongated in polar directions
\citep[e.g.,][]{Schartmann+2014, Wada+2016, Williamson+2020}. Note
that only large dust grains could survive in such polar regions
\citep[e.g.,][]{Tazaki_Ichikawa_2020}. The full picture likely
requires magnetohydrodynamics as well \citep{Blandford+Payne_1982,
  Emmering+1992} given the recent direct detection of $B$-fields along
the equatorial direction of the dusty torus in NGC~1068
\citep{Lopez-Rodriguez+2020}. These measurements of magnetically
aligned dust grains using ALMA polarization at 860~\mic\ support the
picture that magnetic fields up to a few parsecs contribute to the
accretion flow onto active nuclei.

We also compared the emission morphologies to their generative
projected (2D) dust maps. A key result is that the emission morphology
does not simply trace the dust distribution in the system. Optical
depth effects are important at most wavelengths, including in the
MIR. In the FIR and submillimeter regimes the torus is mostly
optically thin, even for a high number of clouds along the LOS. There,
the observed emission does trace the underlying dust distribution.

To quantify morphologies, we advocate the use of moment-based
techniques (e.g., radii of gyration) over more traditional methods
(e.g., half-light radius). Moment-based morphological quantities are
independent of the image brightness. Thus, in our case, maps of the
dust cloud distribution are independent of \No\ and \tv, since these
are just multiplicative factors, and the brightness maps are
independent of the total flux. Furthermore, morphological size
measurements using \emph{radii of gyration} provide size estimates
along both axes independently, making it trivial to measure
elongations. Moment-based quantities can estimate higher-order
morphological quantities such as image asymmetry, skewness, and
kurtosis.

In anticipation of the extremely large telescopes currently planned or
under construction (\GMT, \TMT, \ELT), and to facilitate studies of
image morphology, we have introduced with \HC\ a hypercube of clumpy
torus \emph{images} spanning a wide range of wavelengths and model
parameters. The sampled wavelengths include important NIR, MIR, FIR,
and submillimeter bands. \HC\ also comprises a \emph{software suite}
whose key functionality is to interpolate the images continuously at
any set of model parameters. Another capability of \HC\ is to simulate
several observing modes, e.g., single-dish observations with PSF
convolution\footnote{The pupil images of these telescopes, from which
  PSFs can be computed, are provided with the software and in the
  supplemental materials.}, detector pixelization, noise modeling, and
image deconvolution to recover some of the original signal in the
images. The processing software may also be applied to analyze
different emission models, which need not be limited to AGN tori.

In the companion \papertwo\ we apply the capabilities of the \HC\
software to the \C\ image hypercube and simulate observations of
NGC~1068 with several single-dish telescopes (\JWST, \Keck, \GMT,
\TMT, \ELT) to investigate the spatial resolvability of the NIR to MIR
emission. We also simulate IFU-like observations to analyze the
10-\mic\ silicate emission feature. Finally, we fit directly with the
\C\ images the $K$-band image of NGC~1068 recently published by
\gravity, deriving likely model parameters.

\section{Supplementary Material and Software}
\label{sec:supplements}
\noindent
The hypercubes of \C\ images and cloud distribution maps (as a
function of all model parameters and wavelengths) are provided to the
community through FTP download. See Appendix
\ref{sec:appendix-download}, and the User Manual, for download
instructions.

We also provide the \HC\ software, which comprises user-friendly tools
to operate on the hypercubes, simulate observations, and investigate
morphology. All codes, scripts, and telescope pupil images can be
found in the GitHub repository at
\url{https://github.com/rnikutta/hypercat/} .

The repository also contains the User Manual, several instructional
Jupyter notebooks, and API documentation. The User Manual contains
high-level usage examples and also showcases low-level operations that
can be performed with \HC.

We kindly ask for citation of this paper if the reader decides to make
use of any of the \HC\ data cubes, software functions, scripts, or
telescope pupil images.

\acknowledgments

{We wish to thank Leonard Burtscher, Konrad Tristram, Marko Stalevski,
  Moshe Elitzur, Ric Davies, and Tanio Diaz Santos for illuminating
  discussions on the subjects of this paper.
We are thankful to the referee, whose comments helped improve the manuscript.
R.N. acknowledges early support by FONDECYT grant No. 3140436.
E.L.-R. acknowledges support from the Japanese Society for the Promotion
of Science (JSPS) through award PE17783, the National Observatory of
Japan (NAOJ) at Mitaka, and the Thirty Meter Telescope (\TMT) Office at
NAOJ-Mitaka for providing a space to work and great collaborations
during the short stay in Japan.
K.I. acknowledges support by the Program for Establishing a Consortium
for the Development of Human Resources in Science and Technology,
Japan Science and Technology Agency (JST), and partial support by the
Japan Society for the Promotion of Science (JSPS) KAKENHI (20H01939;
K.~Ichikawa).
C.P. acknowledges support from the NSF grant No. 1616828.
S.F.H. acknowledges support by the EU Horizon 2020 framework program
via the ERC Starting grant DUST-IN-THE-WIND (ERC-2015-StG-677117).
A.A.-H. acknowledges support through grant PGC2018-094671-B-I00
(MCIU/AEI/FEDER,UE). A.A.-H. work was done under project
No. MDM-2017-0737 Unidad de Excelencia ``Mar\'{\i}a de Maeztu'' -
Centro de Astrobiolog\'{\i}a (INTA-CSIC).
R.N., E.L.-R., K.I. are very grateful to NOAO (now part of NSF's
NOIRLab), SOFIA Science Center, and to the Program for Establishing a
Consortium for the Development of Human Resources in Science and
Technology, Japan Science and Technology Agency (JST), for providing
travel grants that made three on-site project meetings possible.
We are grateful to colleagues at several telescope collaborations and
consortia who were willing and able to provide us with the latest
versions of their pupil images. These are, in order of increasing
telescope diameter: Marshall Perrin (\JWST), Andrew Skemer (\Keck),
Warren Skidmore and Christophe Dumas (\TMT), Rebecca Bernstein (\GMT),
and Suzanne Ramsey (\ELT). With their permission we are here
publishing these pupil images as FITS files (see
Section \ref{sec:supplements}, and the project repository
\url{https://github.com/rnikutta/hypercat/}).  }

\facilities{\JWST, \Keck, \VLTI\ (GRAVITY)}

\software{
  \HC\ (this work),
  \texttt{astropy} \citep{astropy:2013,astropy:2018},
  \texttt{h5py} \citep{h5py}, \texttt{matplotlib} \citep{matplotlib},
  \texttt{numpy} \citep{numpy}, \texttt{r2py} \citep{r2py},
  \texttt{scipy} \citep{scipy}, \texttt{urwid} \citep{urwid}}


%
\bibliography{paper}{}
\bibliographystyle{aasjournal}

%

\appendix

\section{Download and Verification of Model Hypercubes}
\label{sec:appendix-download}
\noindent
Readers interested in using the \C\ hypercubes will need to download
one or more \texttt{hdf5} files from
\url{https://www.clumpy.org/images/}, or from the current FTP
location:
\url{ftp://ftp.tuc.noirlab.edu/pub/nikutta/hypercat/}. Table~\ref{table:hdf5files}
lists the available options. The \texttt{*\_all.hdf5.gz} file contains
the image hypercube at all sampled wavelengths. This is the maximally
compressed version of the hdf5 file, which must be uncompressed on the
user's computer system. To reduce the peak storage required on the
target computer, both steps can be executed in one go (all commands in
a single line):
\begin{lstlisting}[name=download,caption={}]
lftp -e 'set net:timeout 10; cat /pub/nikutta/hypercat/hypercat_20200830_all.hdf5.gz; bye' ftp.tuc.noirlab.edu | gunzip > 
  hypercat_20200830_all.hdf5
\end{lstlisting}
The program \texttt{lftp} must be installed on the target system, and
914~GB of space must be available on it (but only 271 GB of compressed
data will be downloaded). The reader should also download the
checksums file
\url{ftp://ftp.tuc.noirlab.edu/pub/nikutta/hypercat/hypercat_20200830.md5}
file and verify the hypercube file:
\begin{lstlisting}[name=md5sum,caption={}]
# this can take 30 minutes even on a modern computer
md5sum --ignore-missing -c hypercat_20200830.md5
hypercat_20200830_all.hdf5: OK

# or on MacOS and BSD variants
md5 hypercat_20200830_all.hdf5
#... and compare the printed hash with the one in the .md5 file
\end{lstlisting}
The download was successful if the computed checksum is identical to
the one in the MD5 sums file. The same procedure applies to the other
\texttt{*hdf5.gz} files analogously (which hold subsets of the
wavelength sampling).

\section{Computation, Storage Requirements, and Organization of HDF5 Model Files}
\label{app:hdf5organization}

\noindent
The \HC\ model images are embedded within the same FOV, corresponding
to the largest sampled \Y\ value, \Ymax. The total number of pixels in
all stored brightness maps is therefore
$N_{\rm pix}^{\rm img} = N_x\, N_y \times \prod_k N_{\theta_k}$, where
$N_{\theta_k}$ is the number of sampled values per model parameter
$\theta_k$, $k \in \{\sig, \iv, \Y, \No, \q, \tv, \lambda\}$, as
compiled in Table~\ref{tab:parameter_sampling}. Thus,
$\prod_k\! N_{\theta_k}$ is the total number of realized parameter
combinations. If full images are stored, the total number of pixels in
the hypercube is
\hbox{$N_{\rm pix}^{\rm img} = \left( 2\,\Ymax\,\etapix + 1 \right)^2
  \times \prod_k N_{\theta_k}$}. If half-images are stored, as is the
case in \HC\ because of image symmetry, then
\hbox{$N_{\rm pix}^{\rm img} = \left( \Ymax\,\etapix + 1 \right)
  \left( 2\,\Ymax\,\etapix + 1 \right) \times \prod_k
  N_{\theta_k}$}. That is, the number of computed pixels grows
quadratically with \Ymax, quadratically with the image resolution per
unit linear size \etapix, and linearly with the sampling of any model
parameter including wavelength. \C\ also computes integrated dust maps
during ray-tracing, and with the parameter sampling of our grid, 4000
such maps were produced (see Section \ref{sec:cloudmap} and
\ref{sec:compstorage}). With half-image ray-tracing and $\etapix = 6$,
the final number of stored pixels (images + cloud maps) is then
$N_{\rm pix}^{\rm tot} = N_{\rm pix}^{\rm img} + N_{\rm pix}^{\rm cld}
\approx 2.45 \times 10^{11}$ pixels, which at single precision (32 bit
= 4 bytes) requires \hbox{$\approx 913$~GB} of storage. We deliver
this hypercube of half-images as a compressed HDF5 file
(\texttt{.hdf5.gz}) of 270~GB size,\footnote{The full hypercube of
  data for this release can be found in HDF5 format, together with MD5
  checksums, at \url{https://www.clumpy.org/images/}} and we have
also produced four smaller HDF5 files with partial wavelength
coverage. Table~\ref{table:hdf5files} lists their properties.

Such file sizes can be handled relatively easily with regard to
network download and possibly even storage on a researcher's personal
workstation, but they are 1-3 orders of magnitude too large to load
the entire set into memory (for instance, for the purpose of
$n$-dimensional interpolation). We must therefore devise routines for
slicing the hypercube along any parameter axes of interest, and for
on-the-fly access to any image map within the envelope of the sampled
parameter space (see User Manual for details). Together with the
hypercubes, we provide functions to transparently extract full-sized
(square) images via $n$-dimensional interpolation on the hypercube and
(automatic) left-right mirroring about the $y$-axis.

The Hierarchical Data Format version 5 (HDF5) is well established for
packaging structured and heterogeneous data sets and metadata, and it
is agnostic about the operating system. We use the Python package
\texttt{h5py} to construct our hypercubes, and the \HC\ software uses
\texttt{h5py} to access said data. HDF implementations in other
languages are readily available.\footnote{E.g., the \texttt{HDF5}
  module in IDL.} The structure of every \HC\ HDF5 file is
simple. Listing~\ref{lst:hdf5} shows the hierarchy tree of the main
HDF5 file released with this paper.
\begin{lstlisting}[name=general,caption={Structure of a \HC\
HDF5 file. Some nonessential attributes omitted for
clarity.},label={lst:hdf5}]
/Nhypercubes : 2
/hypercubenames : ['imgdata', 'clddata']
/imgdata/ : (group)
  /Nparam : 9
  /paramnames : ['sig','i','Y','N','q','tv','wave','x','y']
  /theta : [[15,30,45,60,75],[0,10,20,30,40,50,60,70,80,90],...]
  /hypercubeshape : [5,10,16,12,5,7,25,121,241]
  /hypercube : (9-d array)
/clddata/ : (group)
  /Nparam : 6
  /paramnames : ['sig','i','Y','q','x','y']
  /theta : [[15,30,45,60,75],[0,10,20,30,40,50,60,70,80,90],...]
  /hypercubeshape : [5,10,16,5,121,241]
  /hypercube : (6-d array)
\end{lstlisting}

\noindent
There are \texttt{Nhypercube=2} hypercubes in the file. Their names
are \texttt{imgdata} and \texttt{clddata}, each stored in a separate
HDF group. Group \texttt{imgdata} contains the thermal emission maps,
stored in a 9-dimensional ($\sig,\iv,\Y,\No,\q,\tv,\lambda,x,y$) array
\texttt{imgdata/hypercube}. Several other small data sets of auxiliary
information are also stored inside the group, e.g., the mapping of
parameter values to the hypercube grid vertices (\texttt{theta}). The
hypercube shape is given in \texttt{hypercubeshape}, and every array
axis corresponds to one list in \texttt{theta}, and to one parameter
name in \texttt{paramnames}. \texttt{theta} in the listing above shows
the sampling of the parameters \texttt{sig} and \texttt{i}, which is
the torus angular thickness \sig, and of the viewing angle \iv. Other
parameter samplings are truncated here for brevity. The
\texttt{clddata} group does the same for the 2D maps of cloud number
along the LOS. These maps do not depend on wavelength and cloud
optical depth \tv, and \No\ is just a multiplicative factor (see
Section \ref{sec:cloudmap}); thus, their hypercube is 6-dimensional
($\sig,\iv,\Y,\q,x,y$).

\section{Morphology}
\label{sec:appendix-morphology}
\noindent
Various methods exist to quantify image morphology. One very powerful
approach, due to its generality, is the framework of image
moments. For digitized images $I(x,y)$ with $x$ and $y$ pixel indices,
the $(p+q)$-order \emph{raw} or \emph{geometric} moment is defined as
\begin{equation}
  \label{eq:appendix-momentraw}
  M_{pq} = \sum_x \sum_y I(x,y)\, x^p\, y^q,
\end{equation}
with a standard power basis. Other bases, e.g., orthogonal functions,
Zernike polynomials, etc., are possible and are being used in numerous
applications. For binary images the raw moment $M_{00}$ is the area
(sum) of the one-valued pixels. For grayscale images, in the
astronomical context, it is the total light contained in all pixels.

\begin{figure*}
  \center
  \includegraphics[width=1.\hsize]{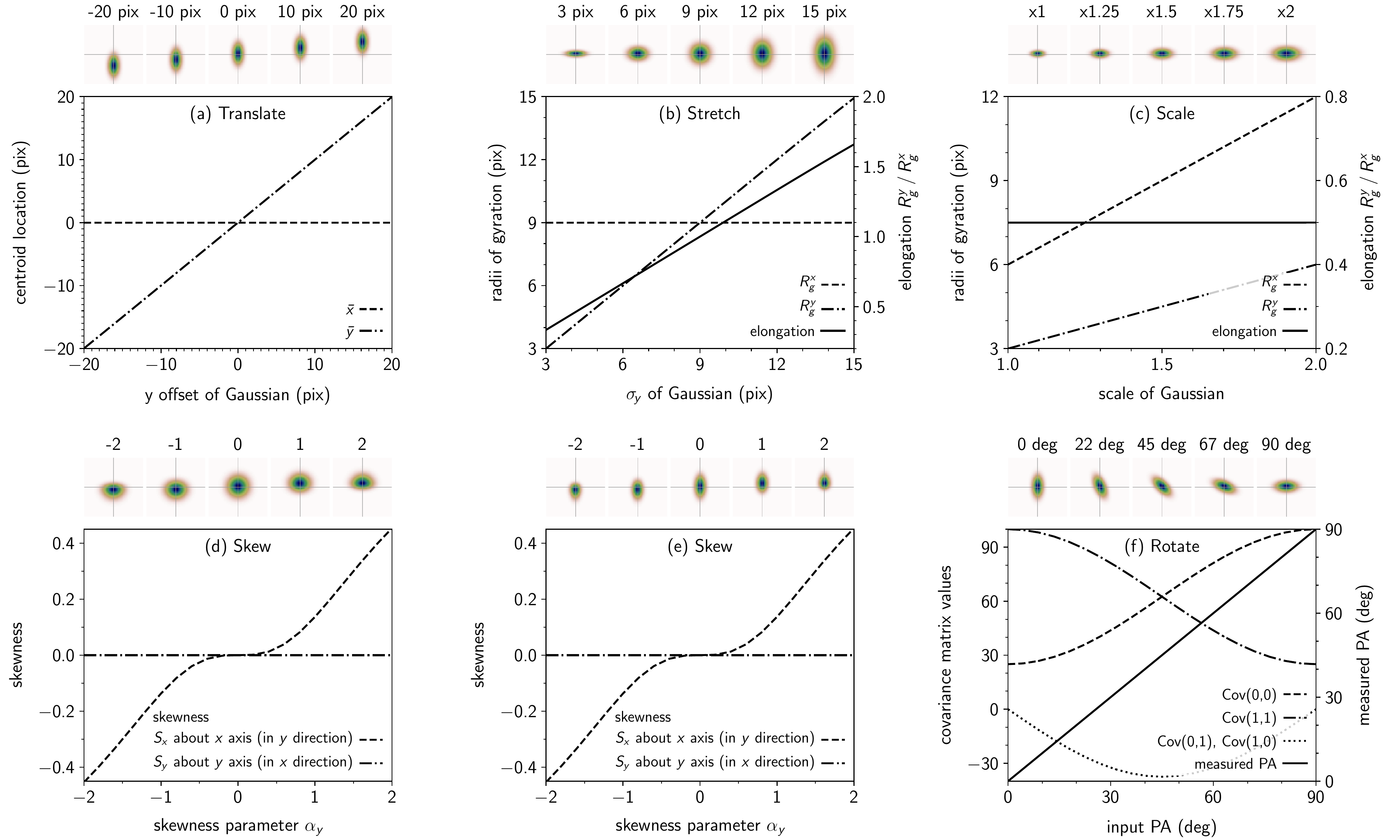}
  \caption{The effects on morphological quantities of various image
    operations, measured on 2D Gaussians. Five small images per
    operation demonstrate the effect. The large panels show the
    measurements as continuous functions. Solid lines use the right
    \yax. (a) Translation, measured via centroid location
    $\bar x, \bar y$. (b) Stretching in the $y$-direction, measured
    via radii of gyration \Rgx, \Rgy. The $y$ elongation of the
    Gaussian is $e = \Rgy\,/\,\Rgx$ (solid line). (c) Scaling of the
    morphology, measured via radii of gyration. The elongation remains
    constant. (d) Skewness about \xax\ (i.e., in the $y$-direction) of
    a circular Gaussian, measured via third-order moments. $\alpha_y$
    controls the skew in the $y$-direction of the 2D skew-normal
    distribution. $\alpha_x$ is fixed at 0, i.e. no skew. (e) Same as
    \emph{(d)} but of a vertically elongated 2D Gaussian. Skewness in
    the $y$-direction is independent of the distribution of pixels in
    the $x$-direction. (f) Rotation of position angle, measured via
    image covariance matrix. The four matrix elements are plotted as
    nonsolid lines. The measured \pa\ (solid line) is observed from
    the image (in degrees east from north, with north up).}
  \label{fig:moment_measures}
\end{figure*}

Figure~\ref{fig:moment_measures} demonstrates the effects of various
image operations (using 2D Gaussians as images) on the measured
moments, from left to right and top to bottom: (a) translation,
(b) stretching, (c) scaling, (d) \& (e) skewing, and (f) rotation.

Panel (a) shows a 2D Gaussian with $\sigma_x=5,\,\sigma_y=10$~pixels
(in a $101 \times 101$ pixel frame) being shifted from $y_0=-20$
through $+20$~pixels, and the corresponding centroid locations
$\bar x$ (not changing) and $\bar y$ (changing).

Panel (b), left axis, shows how \Rgy\ grows when the 2D Gaussian is
being stretched vertically, and in panel (c), left axis, shows how
\Rgx\ and \Rgy\ both increase linearly with the increasing scale of a
2D Gaussian. This linear behavior makes radii of gyration very well
suited for measuring the characteristic sizes of the pixel
distribution.

Panel (b), right axis, also shows how elongation is measured, using a
2D Gaussian as an example whose $x$-size $\sigma_x$ is 9 pixels
always, while $\sigma_y$ is varied from 3 to 15 pixels. That is, the
Gaussian spans a range of aspect ratios from 1:3 to 5:3. The measured
elongation (solid line) grows linearly with $\sigma_y$.

Panels (d) and (e) show how skewness can be measured using
Eqns.~\eqref{eq:skew}. For this demonstration we use the 2D version
(i.e. $k=2$) of the multivariate skew-normal distribution as defined
by \citet{Azzalini_DallaValle_1996}, with density function
\begin{equation}
  \label{eq:mvskewnormal}
  f(z) = 2\, \phi_k(z;\Omega)\,\Phi(\alpha^\mathrm{T} z),\quad (z \in \mathbb{R}^k),
\end{equation}
where $\phi(\cdot)$ is the $k$-dimensional normal density with zero
mean and correlation matrix $\Omega$, $\Phi(\cdot)$ is the $N(0,1)$
distribution function, and $\alpha$ is a $k$-dimensional vector of
skewness coefficients. In our case,
$\alpha = (\alpha_x = 0, \alpha_y)$.\footnote{To compute the
  multivariate skew-normal distributions, we use the \texttt{sn.dmsm}
  implementation in \textsc{R}:
  \url{https://www.rdocumentation.org/packages/sn/versions/1.5-2/topics/dmsn}}
In panel (d) a symmetric 2D Gaussian is modified by changing the
skewness parameter $\alpha_y$. The skewness $S_x$ about the \xax,
i.e., in the $y$-direction, changes as a result, while $S_y$ remains
constant. In panel (e) the same operations are applied to an initially
elongated 2D Gaussian; the resulting $S_x$ curve is identical,
i.e. the skewness about an axis is independent of the distribution of
pixels with respect to the other axis.

\subsection{More on Image Moments}
\label{sec:appendix-moments}
\noindent
Versions of central moments that are invariant to both translation and
\emph{scale} operations are possible by properly normalizing the
central moments
\begin{equation}
  \label{eq:momentcentral}
  \eta_{pq} = \frac{\mu_{pq}}{\mu_{00}^{\left(\displaystyle 1+\frac{p+q}{2}\right)}} \qquad\forall\ p+q\ge2.
\end{equation}
Scale invariance in this context means \emph{similitude}, which is the
resizing of a 2D shape while preserving its proportions. Invariants
simultaneously under translation, scaling, and \emph{rotation} can be
defined as combinations of moments \citep{Hu1962, Reiss1991,
  Flusser_Suk_1993}. For our applications we do not need to concern
ourselves with rotation, since we are always able to set the
positional angle $\pa=0$\degr\ in our images. Some useful observations
on moments:
\begin{enumerate}[label=(\alph*)]
\item Zero-order central and raw moments are identical: $\mu_{00} = M_{00}$.
\item First-order central moments are zero: $\mu_{10} = \mu_{01} = 0$.
\item Second-order central moments $\mu_{20}$ and $\mu_{02}$ measure
  the distribution of mass w.r.t. the image axes, i.e. can be used to
  measure source elongation and orientation (see
  Section \ref{sec:aspect}).
\item Third- and fourth-order central moments can measure higher-order
  image properties such as skew/asymmetry (see
  Section \ref{sec:asymmetry}) and kurtosis/peakedness (see
  Section \ref{sec:appendix-kurtosis}).
\end{enumerate}

\noindent
Note that some of the moment-based functionality is available in the
\texttt{photutils} package for \textsc{Python}. Unfortunately, at the
time of writing, there exist several incompatibilities between
\textsc{Astropy}'s \texttt{models} package, the \texttt{scikit-image}
module, and \texttt{photutils}. For that reason, we rely on our own
moment-based implementation. It is straightforward to use,
is performant, and does not have many dependencies. This
\texttt{morphology} module is part of the \HC\ source code
distribution.

\subsection{Image Elongation via Covariance Matrix}
\label{sec:appendix-covariance}
\noindent

An equivalent method exists to measure the elongation $e$ of a
morphology using the eigenvalues of the image covariance matrix. The
eigenvalues and eigenvectors of the covariance matrix $\Sigma$ of an
image $I$ encode the magnitude and direction of the principal
components of $I$, i.e. the 2D extension of the pixel distribution. In
our case we can always set the position angle to $\pa=0$\degr, i.e.,
we are only interested in elongations along the orthogonal
$\mathbf{e}_x$ and $\mathbf{e}_y$ directions of the image. Thus, the
eigenvalues measure the magnitude of the pixel distribution along the
$x$ and $y$-directions.

The covariance matrix $\Sigma$ can be computed in several ways,
e.g., from second-order central moments
\begin{equation}
  \label{eq:cov}
  \Sigma \equiv \mathrm{Cov}[I(x,y)] = 
  \begin{bmatrix}
    \mu'_{20} & \mu'_{11}\\
    \mu'_{11} & \mu'_{02}
  \end{bmatrix},
\end{equation}
where \hbox{$\mu'_{\rm pq} = \mu_{\rm pq} / \mu_{00}$}. The eigenvalues of
$\Sigma$ can be obtained analytically, or conveniently by computer
packages (e.g., \texttt{linalg.eigvals} in \texttt{numpy}). The
elongation of the intensity distribution in $I(x,y)$ is then the
square root of their ratio, i.e.
\hbox{$e = \sqrt{\lambda_y/\lambda_x}$}. The same elongation can also
be computed using the gyration radii as $e = \Rgy\, /\, \Rgx$. Thus,
$\lambda_x = (\Rgx)^2$ and $\lambda_y = (\Rgy)^2$.

\subsection{Kurtosis}
\label{sec:appendix-kurtosis}
\noindent
Kurtosis is a fourth-order measure that expresses the ``peakedness'' of a
distribution
\begin{equation}
  K_{\!x} = \frac{\mu_{40}}{\mu_{20}^2} - 3, \qquad K_{\!y} = \frac{\mu_{04}}{\mu_{02}^2} - 3.
  \label{eq:kurtosis}
\end{equation}
In the context of image morphology, it can be used to quantify the
compactness of the emission pattern. $K_{x,y}$ are zero for a Gaussian
distribution, greater than zero for distributions more peaked than
Gaussians, and less than zero for flatter distributions.

\subsection{Image Orientation}
\noindent
During our analysis of image morphology in Section~\ref{sec:morphology}
we could set the image orientation to $\pa=0$\degr\ in all
cases. Should, however, the need arise to measure the orientation of
the dominant image component, \HC\ can do this via the covariance
matrix of the image and its eigenvectors:
\begin{lstlisting}[name=orientation,caption={}]
from hypercat import imageops, morphology
cube = ... # load cube as described e.g. in the HYPERCAT User Manual
img = cube(vec) # with parameter vector suitable for cube
rotimg = imageops.rotateImage(img,'42 deg') # PA=42 deg as example
morphology.get_angle(rotimg)
  42.0011  # PA measured from image
\end{lstlisting}

\subsection{Half-light Radius}
\label{sec:appendix-halflightradius}
\noindent
A circular aperture with half-light radius $R_{\rm HL}$ contains half
of the flux in an image $I$, such that \citep[see,
e.g.,][]{Burtscher+2013}
\begin{equation}
  \label{eq:halflightradius}
  \frac{1}{F_{\rm tot}} \int\limits_0^{R_{\rm HL}}\!\!\dif r\, I\, 2 \pi r = \frac{1}{2},
\end{equation}
where $R_{\rm HL}$ is the root of this equation. The solution is not
unique if the circle need not be centered in the image.

\subsection{Gini Coefficient}
\label{sec:appendix-gini}
\begin{figure}
  \center
  \includegraphics[width=0.5\hsize]{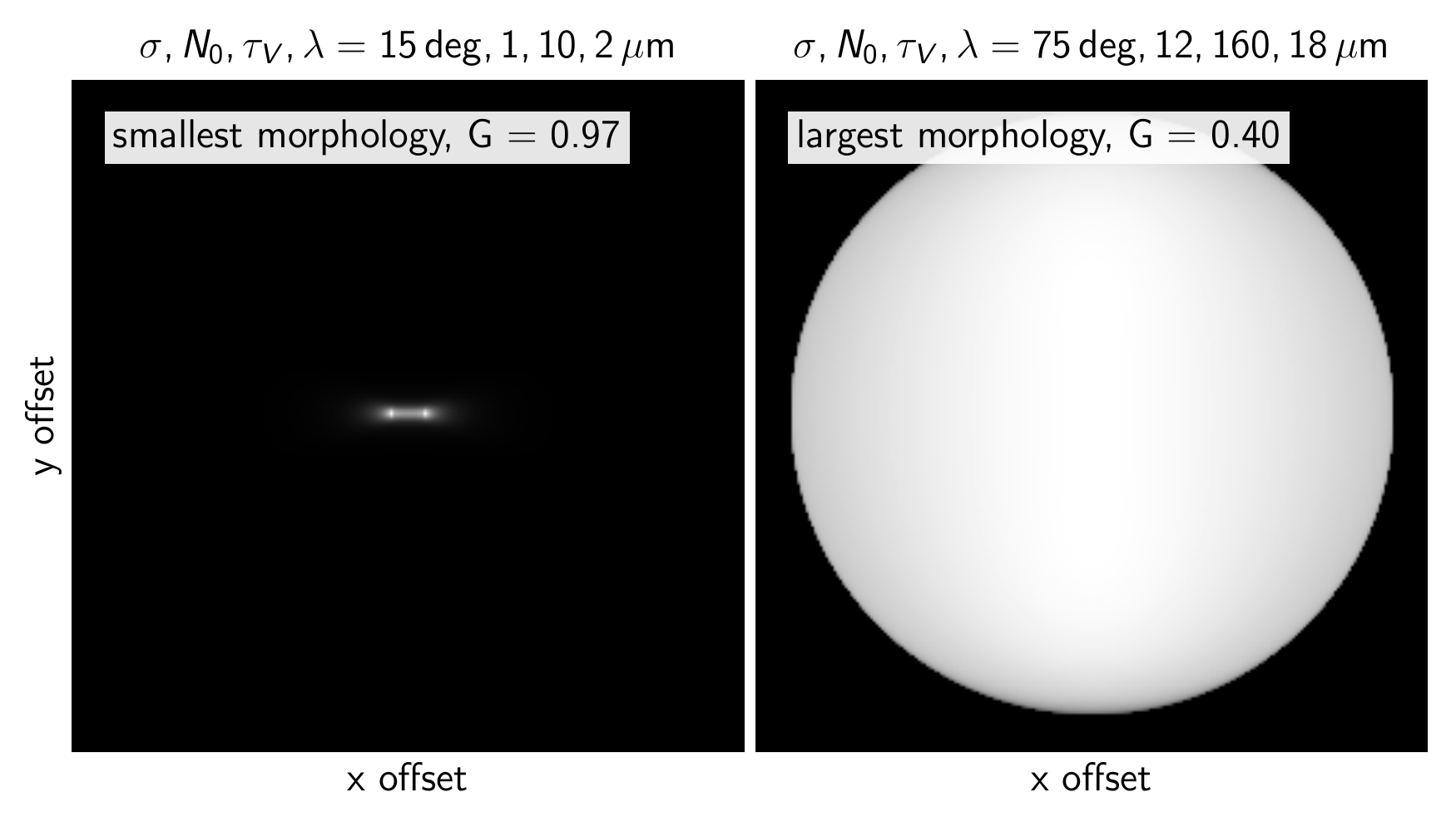}
  \caption{Smallest and largest morphologies within the parameter
    space from Fig~\ref{fig:gini}. Linear scale. $Y=18$, $q=0$, and
    $i=90$\degr\ are identical in both cases. Left: smallest / most
    compact emission map. Gini index $G = 0.97$, elongation
    \hbox{$e = \Rgy / \Rgx = 0.31$}. Right: largest / most extended
    emission map. Gini index $G = 0.4$, elongation $e = 1.0$.}
  \label{fig:minmaxtorus}
\end{figure}

\noindent
Figure \ref{fig:minmaxtorus} shows the smallest and largest
morphologies within the parameter space from Figure \ref{fig:gini}
%

\label{lastpage}
\end{document}